\documentclass[aps,prd,preprint,superscriptaddress,preprintnumbers,nofootinbib]{revtex4-1}
\pdfoutput=1

\usepackage{graphics}
\usepackage[dvips]{graphicx}
\usepackage{mathrsfs}
\usepackage{amssymb}
\usepackage{amsmath}
\usepackage{verbatim}
\usepackage{float}
\usepackage{slashed}
\usepackage{bbm}
\usepackage[dvips,letterpaper,text={6.5in,9in}]{geometry}

\newcommand{\gev}{\text{GeV}}

\newcommand{\tev}{\text{TeV}}
\newcommand{\gu}{\tilde g_u'}
\newcommand{\gd}{\tilde g_d'}
\newcommand{\ra}{\rightarrow}
\newcommand{\re}{\mathrm{Re}}
\newcommand{\im}{\mathrm{Im}}

\usepackage{accents}
\newlength{\dhatheight}

\newcommand{\gsim}{\gtrsim}
\newcommand{\lsim}{\lesssim}

\newcommand{\eg}{{\it e.g.}}
\newcommand{\ie}{{\it i.e.}}
\newcommand{\etc}{{\it etc}}

\begin{document}

\title{Split Dirac Supersymmetry: \\ An Ultraviolet Completion of Higgsino Dark Matter}

\author{Patrick J. Fox}
\affiliation{Theoretical Physics Department, Fermilab, P.O.~Box 500, 
             Batavia, IL 60510, USA}

\author{Graham D. Kribs}
\affiliation{School of Natural Sciences, Institute for Advanced Study, 
             Princeton, NJ 08540}
\affiliation{Department of Physics, University of Oregon,
             Eugene, OR 97403}
             
\author{Adam Martin}
\affiliation{Department of Physics, University of Notre Dame,
             Notre Dame, IN 46556, USA}

\preprint{FERMILAB-PUB-14-124-T} 

\begin{abstract} 

\vspace*{0.5cm} 
Motivated by the observation that the Higgs quartic coupling runs to zero at an 
intermediate scale, we propose a new framework for models of split supersymmetry, in 
which gauginos acquire intermediate scale Dirac masses of $\sim 10^{8-11}$~GeV\@.
Scalar masses arise from one-loop finite contributions as well as 
direct gravity-mediated contributions.  Like split supersymmetry, 
one Higgs doublet is fine-tuned to be light.  The scale at which the 
Dirac gauginos are introduced to make the Higgs 
quartic zero is the same as is necessary for gauge coupling unification.  
Thus, gauge coupling
unification persists (nontrivially, due to adjoint multiplets), 
though with a somewhat higher unification scale
$\gsim 10^{17}$~GeV\@.  The $\mu$-term is naturally at the weak scale, 
and provides an opportunity for experimental verification. 
We present two manifestations of Split Dirac Supersymmetry. 
In the ``Pure Dirac'' model, the lightest Higgsino must decay through $R$-parity 
violating couplings, leading to an array of interesting signals
in colliders.  In the ``Hypercharge Impure'' model, the bino acquires
a Majorana mass that is one-loop suppressed compared with the Dirac gluino 
and wino.  This leads to weak scale Higgsino dark matter whose overall 
mass scale, as well as the mass splitting between the neutral components,  
is naturally generated from the same UV dynamics.  
We outline the challenges to discovering pseudo-Dirac Higgsino dark matter 
in collider and dark matter detection experiments.

\end{abstract}

\maketitle

\section{Introduction}
\label{sec:intro}

In the minimal supersymmetric extension of the Standard Model (MSSM),  
the Higgs quartic coupling is predicted, in terms of a handful of parameters
that determine the tree-level and loop-corrected contributions. 
Now that the LHC has measured the Higgs mass \cite{Aad:2012tfa,Chatrchyan:2012ufa},
and consequently the quartic coupling in the Standard Model, this measurement 
can be used to reverse-engineer the parameters and relevant mass scales
of the supersymmetric theory.  
Scales well above the weak scale are predicted:
$m_{\tilde{t}} \simeq 5$~TeV for $\tan\beta \gg 1$ and $|A_t| \ll m_{\tilde{t}}$
(e.g.~\cite{Arbey:2011ab,Draper:2011aa,Delgado:2012eu,Feng:2013tvd,Draper:2013oza})
and in Split Supersymmetry
\cite{ArkaniHamed:2004fb,Arvanitaki:2004eu,Giudice:2004tc}
$m_{\tilde{t}} \gtrsim 10^{8}$~GeV for $\tan\beta \simeq 1$
\cite{Giudice:2011cg,Hall:2011jd,ArkaniHamed:2012gw,Ibe:2013rpa}.
The long tail to very large superpartner 
masses results from the vanishing of the tree-level quartic coupling 
in the $\tan\beta \ra 1$ limit.  
Reverse-engineering the mass scales of the MSSM 
is unfortunately not very predictive after all.

Supersymmetric models with Dirac gaugino masses, first studied in \cite{Fayet:1978qc,Polchinski:1982an,Hall:1990hq} with more model-building 
explored in  \cite{Fox:2002bu,Nelson:2002ca,Chacko:2004mi,Carpenter:2005tz,Antoniadis:2005em,Nomura:2005rj,Antoniadis:2006uj,Kribs:2007ac,Amigo:2008rc,Benakli:2008pg,Benakli:2009mk,Benakli:2010gi,Kribs:2010md,Abel:2011dc,Davies:2011mp,Benakli:2011kz,Kumar:2011np,Frugiuele:2011mh,Itoyama:2011zi,Frugiuele:2012pe}, predict the 
Higgs quartic coupling to vanish once the gauginos and their 
scalar counterparts are integrated out\footnote{Assuming 
just the dimension-5 supersoft operator, more on this in 
Sec.~\ref{sec:toolkit}.} \cite{Fox:2002bu}. 
This is an improvement on the MSSM, in so far as there is a 
single prediction for the scale of supersymmetry breaking masses.  
Reverse-engineering this scale, and one finds $M_{\rm D} \sim 10^{11}$~GeV, 
where $\lambda_h(M_D) \simeq 0$ 
(for example, 
\cite{Casas:1994qy,Casas:1996aq,Isidori:2001bm,Ellis:2009tp,EliasMiro:2011aa,Bezrukov:2012sa,Degrassi:2012ry,Buttazzo:2013uya}). 
This is akin to the original Split Supersymmetry models 
\cite{ArkaniHamed:2004fb,Giudice:2004tc,Arvanitaki:2004eu}, 
except that both gauginos and scalars are expected to be within an order of
magnitude of this large intermediate scale.  Unlike Split Supersymmetry
models, however, there are negligible corrections to the running
of the Standard Model quartic coupling for scales below $\ll M_D$
(and hence the difference between the upper bound of $\sim 10^8$~GeV
in Split Supersymmetry models with light gauginos \cite{Giudice:2011cg} 
from $\sim 10^{11}$~GeV in Split Dirac Supersymmetry models). 
Other recent versions of intermediate scale supersymmetry include
\cite{Hebecker:2012qp,Giudice:2012zp,Ibanez:2012zg,Unwin:2012fj,Ibanez:2013gf,Hebecker:2013lha,Hall:2013eko,Hall:2014vga}. 

This is an idealized scenario.  In practice, there are additional 
contributions to the Higgs quartic coupling even in models 
with dominantly Dirac gaugino masses.  For one, anomaly mediation 
provides an irreducible Majorana contribution to gaugino masses
as well as a separate contribution to the adjoint scalar masses,
the latter causing corrections to the pure Dirac prediction of a  
vanishing quartic coupling.  The size of Majorana masses is naturally  
loop-suppressed compared with the Dirac gaugino masses, for example 
in models with gravity-mediation \cite{Kribs:2010md}. 
This leads to a very small contribution to the quartic coupling.  
Another contribution arises from the dimension-6 (so called ``lemon-twist'')
operator~\cite{Fox:2002bu} 
in the superpotential, $W' W' {\rm tr} \Phi^2/M^2$, that results in shifts 
of the masses of scalar components of the adjoint superfields.  
The mass shifts cause an incomplete cancellation of the quartic,
though it is controllable within the order one differences between 
the coefficients of these operators and $\tan\beta$. 
Yet another contribution is a supersymmetric mass for the adjoint
superfield, which shifts both the gaugino masses as well as the 
scalar masses, the latter causing corrections to the pure Dirac 
prediction of a vanishing quartic coupling as before. 
Finally, the superpotential operator that generates Dirac gaugino masses
may not exist for all of the gauge groups.
In the Hypercharge Impure model that we discuss below, 
there is no singlet partner for the bino, and thus, the bino does 
not acquire a Dirac mass.  As a consequence,
the bino acquires a loop-suppressed Majorana mass from anomaly mediation,
and regenerates a small Higgs quartic coupling, 
$\lambda_h \simeq (g'^2 \cos^2 2\beta)/4$.
This implies a restricted range of intermediate scales for the supersymmetry
breaking masses is predicted, between $10^8$ to $10^{10}$~GeV, corresponding to 
between $\tan\beta \gg 1$ to $\tan\beta \simeq 1$.  

The $\mu$ parameter could be small or near the intermediate scale, 
depending on whether a ``bare'' $U(1)_{PQ}$-breaking mass, 
$\int d^4\theta \, H_u H_d$, is permitted \cite{ArkaniHamed:2012gw}. 
As a chiral K\"ahler operator, it is technically natural to omit it,
which we do.  Thus, we consider $B_\mu$ and $\mu$ generated through higher 
dimensional operators after supersymmetry is broken.  
K\"ahler operators at dimension-6 can lead to both 
$B_\mu$ and $\mu$ from $D$-term and $F$-term contributions.  If there are no 
singlets in the hidden sector, which is consistent with the gauginos not acquiring 
Majorana masses (except through anomaly-mediation), 
the leading operator to generate $\mu$ is 
$\int d^4\theta \, {W'}^\dagger {W'}^\dagger H_u H_d/\Lambda^3$, 
which is dimension-7, and thus suppressed relative to the intermediate scale.  
$B_\mu$ can arise through dimension-6 operator in the superpotential 
$\int d^2\theta \, W'W' H_u H_d/\Lambda^2$, 
whose coefficient is set by doing one fine-tuning to get one Higgs doublet light.
Given $B_\mu$, as well as anomaly-mediated Majorana contributions to the 
gaugino masses, both $U(1)_{PQ}$ and $U(1)_R$ are broken in the visible sector
near the intermediate scale, 
and thus there is also a one-loop radiative contribution to $\mu$
\cite{Giudice:2004tc,Arvanitaki:2004eu}.
In the Hypercharge Impure model, this one-loop radiative contribution 
provides the dominant contribution to $\mu$, analogous to one version
of Spread Supersymmetry \cite{Hall:2011jd}, as we will see. 

Remarkably, gauge coupling unification persists when $\mu \sim$~weak scale
with a Dirac gluino and Dirac wino at an intermediate scale.
Gauge coupling unification with intermediate scale Dirac gauginos
has been studied before \cite{Antoniadis:2005em}, and unification occurs
with fairly good accuracy even when light Higgsinos are the only 
new physics affecting gauge coupling running 
\cite{ArkaniHamed:2005yv,Mahbubani:2005pt}. 
In the models we consider, given a weak scale $\mu$ parameter, 
the leading difference at one-loop from the MSSM is the \emph{scale}
of the Dirac gaugino masses and the additional \emph{degrees of freedom} 
due to the additional adjoint chiral superfields.  Since the
degrees of freedom are proportional to the appropriate quadratic 
Casimir of the group [$N$ for $SU(N)$], there is some common 
Dirac gaugino mass scale where gauge coupling unification \emph{must occur}.  
Remarkably, we find $M_D \simeq 10^{11}$~GeV, which is essentially 
the same scale where $\lambda_h(M_D) \simeq 0$.  
The additional degrees of freedom (Dirac fermion partner and complex
scalar in the adjoint representation) accelerate the RG evolution of the
gauge couplings between the intermediate scale to the unification scale 
in such a way as to exactly compensate for the lack of Majorana gauginos 
in the RG evolution between the weak scale and the intermediate scale.
This is discussed in Sec.~\ref{sec:simpleunify}.

\begin{figure}[t]
\centering
\includegraphics[width=0.6\textwidth]{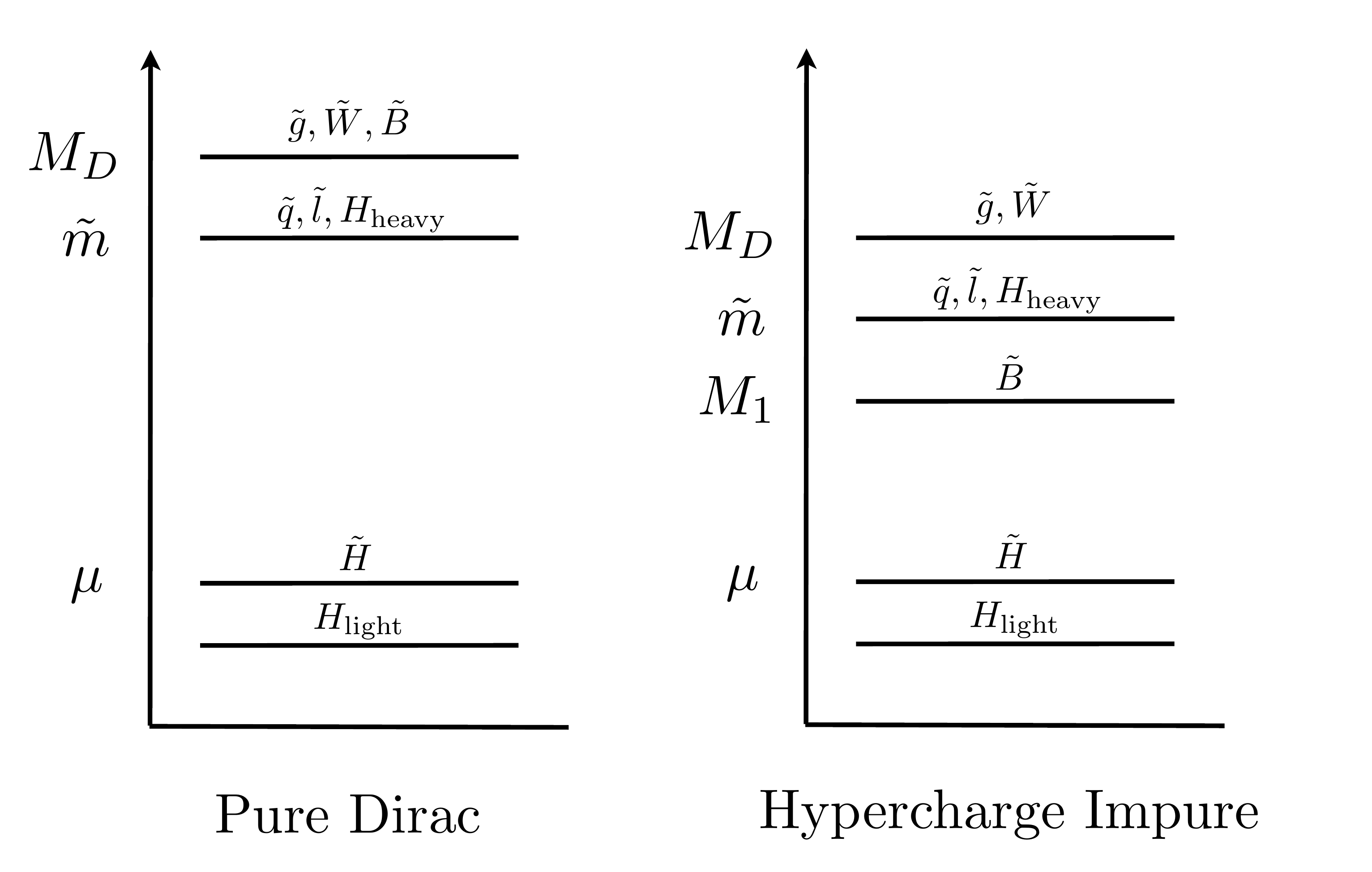}
\caption{Sketch of the mass spectrum of the two split Dirac supersymmetry models
considered in this paper:  Pure Dirac (all gauginos acquire Dirac masses) 
and Hypercharge Impure (the gluino and wino acquire Dirac masses, the bino
acquires a Majorana mass).}
\label{fig:spectrum}
\end{figure}

The outline of the paper is as follows.  We first present the ``toolkit''
for Split Dirac Supersymmetry models in Sec.~\ref{sec:toolkit}.
This includes the variety of operators and contributions to the
soft masses and $\mu$ parameter in the theory.
We demonstrate gauge coupling unification is successful at one-loop
in Sec.~\ref{sec:simpleunify}.  
Gauge coupling unification, however, is not directly affected by the 
character of the bino, i.e., whether there is (or is not) a pure singlet
superfield for it to acquire a Dirac mass. 
This leads to two distinct models within the larger framework of
Split Dirac Supersymmetry:
\begin{itemize}
\item ``Pure Dirac'' model (Sec.~\ref{sec:puredir}):  The gluino, wino, 
and bino acquire Dirac masses.  In this model, the Higgs quartic coupling 
vanishes at the intermediate scale, and thus predicts the largest scale 
for the Dirac gauginos.
The Higgsino mass is small, arising from a dimension-7 operator 
as well as a suppressed radiative contribution.   The neutral Higgsinos 
are highly degenerate, $\Delta m_{\chi} \ll \text{keV}$, 
forming a nearly pure Dirac fermion with an unsuppressed $Z$ coupling,
and are ruled out as a dark matter candidate.  $R$-parity violation is
introduced, and we demonstrate the various decay modes that are 
possible for the lightest Higgsino.
\item ``Hypercharge Impure'' model (Sec.~\ref{sec:mbino}):
The gluino and wino acquire Dirac masses,
while the bino acquires a Majorana mass from anomaly-mediation, 
making it lighter than the other gauginos.  In this scenario, 
a small quartic coupling may be regenerated, depending on
$\tan\beta$ (which in turn depends on the relative hierarchy between 
$B_\mu$ and $m_{H_u}^2,m_{H_d}^2$).  Generally, a slightly lower
scale for $M_D \sim 10^{8} \ra 10^{9}$~GeV results, causing 
$M_1 \sim 10^{6} \ra 10^{7}$~GeV\@.  This large bino mass has the
feature of generating the scale of $\mu$ \emph{and} the mass splitting
$m_{\tilde{\chi}_2} -  m_{\tilde{\chi}_1} \simeq M_Z^2 \sin^2\theta_W/M_1$
to make the lightest Higgsino a perfect WIMP candidate for dark matter.
\end{itemize}
The mass spectra associated with each of these models is shown in
Fig.~\ref{fig:spectrum}.
Finally, we conclude with a discussion in Sec.~\ref{sec:discussion}.

\section{Toolkit for Split Dirac Supersymmetric Models}
\label{sec:toolkit}

Split Dirac supersymmetry is a general framework for considering
a new class of split supersymmetry models.  In this section, we provide 
an overview of the operators leading to contributions to the supersymmetry 
breaking and preserving parameters in the (Dirac extended) MSSM\@.  
This serves as a ``toolkit'' with which split Dirac supersymmetry
model enthusiasts can build interesting models.
We use the results of the toolkit to construct the two models 
that serve as the focus of this paper in Secs.~\ref{sec:puredir}
and \ref{sec:mbino}. 

\subsection{No singlets in the hidden sector}

Majorana gaugino masses arise when total gauge singlets in the
hidden sector, $S$, acquire supersymmetry breaking vevs for their
$F$-components, $S = F \theta^2$.  The usual dimension-5 operator
that leads to Majorana gauginos is
$\int d^2\theta \, S \, W_\alpha W^\alpha/\Lambda$.  
While it is always technically natural to omit these contributions, 
if there no singlets in the hidden sector, this operator is 
simply forbidden.  In addition, the absence of hidden sector singlets 
also means the usual dimension-5 operator 
in the K\"ahler potential that generates $\mu$, 
$\int d^4 \theta \, S^\dagger H_u H_d/\Lambda$, is forbidden. 
Hidden sectors without singlets are well known,
for example $SU(4) \times U(1)$ \cite{Dine:1995ag}. In the absence of hidden sector singlets,
gauginos can acquire Dirac masses through 
$D$-type expectation values, as explained below, as well as anomaly-mediated Majorana masses.  The $\mu$ term can arise through higher dimensional
operators, or through radiative corrections, as we will see.

\subsection{Dirac Gaugino Masses}
 
A Dirac gaugino mass for one or more gauge groups of the Standard Model
arises once the MSSM is extended with an additional superfield $\Phi_k$ 
in the adjoint representation of the appropriate gauge group, 
$k=1,2,3$ for $U(1)_Y$,\,$SU(2)_L$,\,$SU(3)_c$.  
The Dirac mass is generated through the operator
\begin{equation}
\mathcal L \supset \lambda_k \int\, d^2\theta \sqrt 2\, 
\frac{\mathcal {\mathbf W'}_{\alpha}\, \mathcal W^{k,\alpha}_a \, 
                \Phi_{k,a} }{\Lambda} + h.c., 
\label{eq:massop}
\end{equation}
where $\mathcal {\mathbf W'}_{\alpha} = \theta_{\alpha} \mathbf{D}$ is a 
spurion for supersymmetry breaking, $\mathcal W^{k,\alpha}_a $ is the 
gauge superfield for the appropriate SM gauge group, and $\Lambda$ is the scale 
where supersymmetry breaking is mediated to the visible sector. 
The labels $\alpha$ and $a$ are spinor and gauge indices, respectively. 
Inserting the $D$-term expectation value, the operator gives
\begin{equation}
\mathcal L \supset -M_{D,k} \,
  \left( \lambda_a \psi_a + 2\,\sqrt 2 D_a\, \text{Re}(A_a) \right) + h.c. \, ,
\label{eq:diracmass} 
\end{equation}
where $A_a$ is the complex scalar of the supermultiplet $\Phi_a$.
This term marries the gaugino $\lambda_a$ with a fermion in the 
adjoint representation $\psi_a$ with a Dirac mass
$M_{D,k} \equiv \lambda_k {\bf D}/\Lambda$.  

\subsection{Higgs quartic coupling at dimension-5}

The tree-level quartic coupling for Higgs boson arises from the $D$-terms.  
The new ingredient from the dimension 5 operator of Eq.~(\ref{eq:diracmass}), is the term
$2 \sqrt{2} M_D D_a \re(A_a)$ in addition to the term $-D_a^2/2$
from the gauge kinetic terms in the superpotential.  
Solving for the $D$-term through its equation of motion gives
\begin{equation}
D_a = - 2 \sqrt{2} M_D \re(A_a) + g_a^2 \sum_i \phi_i^* t^a \phi_i \, .
\end{equation}
Substituting this back into Lagrangian,
\begin{eqnarray}
\frac{1}{2} \left( 2 \sqrt{2} M_D \re(A_a) 
                   + g_a^2 \sum_i \phi_i^* t^a \phi_i \right)^2 \, ,
\label{eq:quadratic}
\end{eqnarray}
we find the usual Higgs quartic coupling, a mass for $\re(A_a)$, and a cross-term.
Once $\re(A_a)$ is integrated out at $\simeq M_D$, no quartic couplings 
proportional to gauge couplings remain.  Hence, the tree-level Higgs 
quartic coupling vanishes.

\subsection{Higgs quartic coupling at dimension-6}

There are additional contributions to the quartic coupling. 
Using just $D$-terms, at dimension-6 one can write the lemon-twist 
operator
\begin{equation}
\frac{\lambda_{\rm lt}}{2} 
  \int d^2\theta\, \frac{\mathbf{W'_{\alpha}W'^{\alpha}}}{\Lambda^2} \,
  \text{tr}(\Phi_a\, \Phi_a) + h.c. \, .
\label{eq:lemontwist1}
\end{equation}
This superpotential term gives masses to both $\re(A_a)$ and $\im(A_a)$ 
scalar components of $\Phi_a$, but with opposite sign. 
This additional mass term for $\re(A_a)$ disrupts the quadratic form,
Eq.~(\ref{eq:quadratic}), and thus can re-introduce a partial 
quartic coupling for the Higgs boson.  The size of the quartic
depends on the relative size of the operator coefficients\footnote{Throughout 
this paper, we use the normalization convention 
$V(H) \supset \frac{\lambda_h}{2}(H^{\dag}H)^2$ 
for the Higgs quartic.}, 
\begin{eqnarray}
\Delta \lambda_{h,{\rm tree}} &=& 
    \frac{1}{4} \cos^2 2\beta 
    \left( 
    \frac{\lambda_{\rm lt} g^2}{4 \lambda_2^2 + \lambda_{\rm lt}} + 
    \frac{\lambda_{\rm lt} {g'}^2}{4 \lambda_1^2 + \lambda_{\rm lt}} 
    \right) \, .
\label{eq:lemontwist2}
\end{eqnarray}
In many UV completions this operator is generated at the same order as the operator of 
Eq.~(\ref{eq:massop}), and thus is too large.  However, solutions to this problem have been proposed~\cite{Csaki:2013fla}.
It should also be noted that it is technically natural to omit this contribution 
from the superpotential, so its absence need not require 
tuning coefficients.  It is also true that a modest hierarchy
between the dimension-5 coefficient and the dimension-6 coefficient
will also render this contribution to the quartic coupling to be
negligible.

\subsection{Majorana gaugino masses}

In the absence of singlets in the hidden sector, Majorana gaugino 
masses arise from anomaly-mediation. 
Placed in the context of supergravity and tuning away the 
cosmological constant, supersymmetry breaking generates a 
gravitino mass at least of order
\begin{equation}
m_{3/2} \sim \frac{\mathbf D}{\sqrt 3 M_{pl}} \, .
\end{equation}
(Here we assume the $D$-term dominates the supersymmetry breaking
in the hidden sector.)
The anomaly-mediated contribution to the Majorana gaugino masses is 
\cite{Randall:1998uk,Giudice:1998xp}
\begin{eqnarray}
\tilde{M}_k &=& \frac{\beta_k}{g_k} m_{3/2} \, ,
\label{eq:amsbgaugino}
\end{eqnarray}
where $\beta_k$ are the gauge coupling beta-functions given in
Appendix~\ref{sec:RGE}. 
Comparing the size of the Dirac and Majorana gaugino masses, we find
\begin{eqnarray}
\frac{\tilde{M}_k}{M_{D,k}} &=& \frac{\beta_k}{g_k \lambda_k} 
\frac{\Lambda}{M_{pl}} \, . 
\label{eq:gauginoratio}
\end{eqnarray}
We see that the Majorana gaugino masses are suppressed by at least a loop-factor 
times gauge coupling squared relative to the Dirac gaugino masses.
Further suppression is possible if the mediation scale is below
the Planck scale.  The Majorana mass splits the Dirac gaugino state into 
two Majorana states -- though the loop suppression from
Eq.~(\ref{eq:gauginoratio}) implies that the splitting between the states 
is small and the gauginos are more accurately described as pseudo-Dirac.
Pseudo-Dirac gauginos do not in themselves change the argument 
about the vanishing of the quartic coupling.  In anomaly-mediation,
the gaugino masses are also accompanied by scalar mass squareds
that are two-loop suppressed relative to the gravitino mass, 
but this leads to a very small correction for the Majorana
masses given in Eq.~(\ref{eq:gauginoratio}).

\subsection{Higgs quartic coupling with supersymmetric masses for the adjoints}

Supersymmetric masses for the adjoint fields can be generated 
through the operator
\begin{equation}
\frac{\lambda_{\rm adj}}{2} 
  \int d^4\theta\, \left(
  \frac{\mathbf{{W'}_{\alpha}^\dagger {W'}^{\alpha\dagger}}}{\Lambda^3}\,
  \text{tr}(\Phi_a\, \Phi_a) + h.c. \right) \, ,
\label{eq:adjointmass}
\end{equation}
that gives a very small supersymmetric contribution to the masses of 
the adjoint fields, $M_{\rm adj} \equiv \lambda_{\rm adj} {\bf D}^2/\Lambda^3$.
In principle this contribution modifies the quartic coupling \cite{Fox:2002bu} 
\begin{eqnarray}
\Delta \lambda_{h,{\rm tree}} &=& 
    \frac{1}{4} \cos^2 2\beta 
    \left( 
    \frac{g^2 M_{\mathrm{adj},2}^2}{M_{\mathrm{adj},2}^2 + 4 M_{D,2}^2} + 
    \frac{{g'}^2 M_{\mathrm{adj},1}^2}{M_{\mathrm{adj},1}^2 + 4 M_{D,1}^2} \right).
\label{eq:adjointquartic}
\end{eqnarray}
Given that $M_{\mathrm{adj}} \sim M_D^2/\Lambda$, this leads
to a negligible correction.  If however ``bare'' supersymmetric masses 
for the adjoints were present in the superpotential, 
$\mathcal{O}(1) \int d^2\theta \, M_{\mathrm{adj}} \, 
\mathrm{tr}( \Phi_a \Phi_a  ) + h.c.$,
independent of supersymmetry breaking, with masses of order or exceeding the
Dirac masses, then a partial quartic is recovered. 
For example, in the Hypercharge Impure model detailed in 
Sec.~\ref{sec:mbino}, the bino does 
not acquire a Dirac mass.  This could occur even with the existence
of Eq.~(\ref{eq:massop}) with a bino superfield partner (a total gauge singlet), 
if the mass $M_{\mathrm{adj},1} \gg M_{D,1}$, so that 
$\lambda_h = g'^2 \cos^2 2\beta/4$ from Eq.~(\ref{eq:adjointquartic}).

\subsection{$\mu$ and $B_\mu$ term from $D$-terms}

Using just $D$-type spurions, both $U(1)_{PQ}$ and $U(1)_R$ can be violated through higher dimensional operators. As a result, both $\mu$ and $B_\mu$ can be generated.
  The leading contribution to $\mu$ 
is from
\begin{equation}
\int d^4 \theta \, 
  \frac{\mathbf{W'^{\dag}_{\alpha}W'^{\alpha\dagger} }\, H_u\, H_d}{\Lambda^3} 
  = \int d^2\theta \, \frac{D^2}{\Lambda^3}\, H_u\, H_d 
  = \int d^2\theta \, \frac{M^2_D}{\Lambda}\, H_u\, H_d
\label{eq:mudim7}
\end{equation}
that gives $\mu \sim \tev$ when $M_D \sim 10^{11}\, \gev$ 
and $\Lambda = M_{Pl}$.  Notice also that once 
$M_D \lsim 10^{10}$~GeV (for $\Lambda = M_{Pl}$), 
this contribution becomes too small to give a 
large enough $\mu$ to evade direct collider constraints
on Higgsinos.  We will refer to this $\mu$-term contribution 
as the ``primordial'' $\mu$.

The leading contribution to $B_\mu$ arises from the superpotential
operator
\begin{equation}
\lambda_{B_\mu} \int d^2 \theta\, 
  \frac{\mathbf{W'_{\alpha}W'^{\alpha} }\, H_u\, H_d}{\Lambda^2} 
  = \lambda_{B_\mu} \frac{D^2}{\Lambda^2}\, \tilde{H}_u \, \tilde{H}_d \, .
\label{eq:genBmu}
\end{equation}
Notice that $B_\mu$ is parametrically of the same size as the
Dirac gaugino mass found in Eq.~(\ref{eq:diracmass}).

\subsection{Radiative generation of $\mu$}
\label{sec:radiativemu}

The global symmetries $U(1)_{\rm PQ}$ and $U(1)_R$ are broken
by the $B_\mu$ term and Majorana gaugino masses.  In a model \emph{without}  Dirac mass terms for the bino and wino,  
this implies $\mu$ can be radiatively generated \cite{Giudice:2004tc} 
through the renormalization group equation,\footnote{This result 
includes one minor correction to the RGE for $\mu$ given in 
Ref.~\cite{Giudice:2004tc}.  The correct expression involves
the \emph{complex conjugate} of the gaugino mass, such that 
the reparameterization-invariant phases
${\rm arg}(\tilde{g}_u'^* \tilde{g}_d'^* \mu M_1)$ 
and ${\rm arg}(\tilde{g}_u^* \tilde{g}_d^* \mu M_2)$ 
are not generated if there is no 
primordial contribution to $\mu$.}
\begin{equation}
(4\pi)^2\, \frac{d\mu}{dt} = 
\tilde{g}'_u\,\tilde{g}'_d\,  M^*_1 
+ 3 \tilde{g}_u\,\tilde{g}_d\,  M^*_2 
+ \frac{1}{4}\mu \left[ - 18 \left( \frac{g_1^2}{5} + g_2^2 \right)
                        + 3 (\tilde{g}_u^2 + \tilde{g}_d^2) 
                        + \tilde{g}^{\prime 2}_u + \tilde{g}^{\prime 2}_d \right] \, . 
\label{eq:genmu2}
\end{equation}
The $\tilde{g}'_{u,d}\, (\tilde g_{u,d})$ couplings are the strengths of the up or down-type
Higgsino-Higgs-bino (Higgsino-Higgs-wino) Yukawa couplings. At the scale of supersymmetry 
breaking $M_D$, these Yukawa couplings are related, at tree level, to the gauge couplings as 
$\tilde{g}'_u(M_D) = g'\,\sin{\beta}$, $\tilde{g}'_d(M_D) = g'\,\cos{\beta}$, etc.
However, below $M_D$, the theory is no longer supersymmetric so the RGE
for the Higgsino-Higgs-bino Yukawa couplings is no longer the same as the 
RGE for the gauge couplings.  The RGE for $\mu$ is proportional to 
$\sin({2\beta})$, which vanishes in the limit $\tan\beta \to \infty$ (or $0$). 
This follows because, in this limit, $B_\mu \propto \sin({2\beta}) \ra 0$, 
and hence $U(1)_{PQ}$ symmetry is restored \cite{Giudice:2004tc}.

If however both the bino and wino acquire Dirac masses, the only source of $U(1)_R$
 breaking is the small anomaly-mediated Majorana gaugino mass. Therefore, the RGE in Eq.~(\ref{eq:genmu2}) only applies between the two narrowly split pseudo-Dirac states (between $M_{D,1} \pm \tilde M_1$). As a result, the radiatively generated $\mu$ is highly suppressed. We will see examples of both $M_{D,k} = 0$ and $M_{D,k} \ne 0$ in the models discussed in the Sec.~\ref{sec:mbino}, \ref{sec:puredir}.

\subsection{One-loop finite contributions to scalar masses}

Supersymmetry breaking through $D$-terms is known as 
Supersoft Supersymmetry Breaking \cite{Fox:2002bu} due to the 
finite soft scalar (mass)$^2$ that are induced for the scalars
of the MSSM\@.  The contributions were computed in \cite{Fox:2002bu} to be,
\begin{eqnarray}
\tilde{m}^2 &=& \sum_k \frac{C_k(r) \alpha_k M_{D,k}^2}{\pi} 
                \log \frac{\tilde{m}_{r,k}^2}{M_{D,k}^2} \, .
\label{eq:finitescalar}
\end{eqnarray}
Here $\tilde{m}_{r,k}$ is the scalar mass for the real part of 
the adjoint field, given by $2 M_{D,k}$ in the absence of 
additional contributions from $F$-terms to the scalar masses
(see next subsection).

\subsection{\bf F-term contributions to scalar masses}

Supersymmetry breaking hidden sectors with $D$-term spurions 
(which was utilized above to generate the Dirac gaugino mass) 
generically have spurions, $X$, that transform under the hidden sector group (i.e. non-singlets),
and acquire $F$-terms (e.g., see \cite{Dine:1995ag}). 
The only gauge invariant
combination of the hidden sector spurions ${\bf X}$ that get 
$F$-type expectation values must involve powers of ${\bf X}^\dagger {\bf X}$.  
This implies mass terms for scalars
\begin{equation}
\kappa_i \int d^4 \theta \, \frac{\mathbf{X^{\dag}X}}{\Lambda^2}\, 
  \phi^{\dag}_i\, \phi_i,
\label{eq:fterm}
\end{equation}
as well as a contribution the the $B_\mu$ term,
\begin{equation}
\kappa_{B_\mu} \int d^4 \theta \, \frac{\mathbf{X^{\dag}X}}{\Lambda^2} \, H_u H_d \, ,
\label{eq:Bmufterm}
\end{equation}
are generically present.  These operators give contributions
$|F|^2/\Lambda^2$ to the scalar mass squareds as well as $B_\mu$.

\subsection{\bf Fine-tuning to get one light Higgs doublet}

In split supersymmetry models, fine-tuning in the scalar mass squared
parameters of the Higgs mass matrix is needed such that one doublet 
gets a small, negative mass squared, causing electroweak symmetry breaking
\cite{ArkaniHamed:2004fb,Arvanitaki:2004eu,Giudice:2004tc}
(see also \cite{Hall:2011jd,Ibe:2012hu,Arvanitaki:2012ps,ArkaniHamed:2012gw}). 
In the MSSM, the Higgs mass matrix is 
\begin{eqnarray}
\mathcal{M_H} &=& \left( \begin{array}{cc} m_{H_u}^2 & B_\mu \\
                                           B_\mu     & m_{H_d}^2 \end{array} \right)~,
\end{eqnarray}
where the entries in the mass matrix include all of the supersymmetry
breaking contributions from $D$-terms and $F$-terms described above.
(We have neglected the tiny contribution $|\mu|^2 \ll |m_{H_u}^2|$,$|m_{H_d}^2|$ 
to the diagonal entries.)
Since the Dirac gauginos induce large positive one-loop finite 
contributions to $m_{H_u}^2$ and $m_{H_d}^2$, we assume any 
additional contributions from $F$-terms do not cause these mass-squareds
to go negative.  Electroweak symmetry breaking at the weak scale 
requires one small negative eigenvalue, and hence 
${\rm Det}[\mathcal{M}_H] = m_{H_u}^2 m_{H_d}^2 - B_\mu^2 < 0$.
The light (negative) eigenvalue is
\begin{eqnarray}
m_{H_{\rm light}}^2 \simeq \frac{{\rm Det}[\mathcal{M}_H]}{{\rm Tr}[\mathcal{M}_H]}
  &=& \frac{m_{H_u}^2 m_{H_d}^2 - B_\mu^2}{m_{H_u}^2 + m_{H_d}^2} \nonumber \\
  &=& \frac{1}{1+\tan^2\beta} 
      \left[ m_{H_u}^2 \tan^2\beta - \frac{B_\mu^2}{m_{H_u}^2} \right] \, ,
\end{eqnarray}
where $\tan\beta$ is determined by
\begin{equation}
\tan\beta \simeq \sqrt{\frac{m_{H_d}^2}{m_{H_u}^2}} \, ,
\end{equation}
up to corrections of order $m_{H_{\rm light}}^2/(m_{H_u} m_{H_d})$. 
Clearly we must fine-tune $B_\mu^2$ to be slightly larger than 
$m_{H_u}^2 m_{H_d}^2$ to obtain a small negative mass-squared
eigenvalue.  

It is interesting to compare the size of $B_\mu$ to the one-loop
(finite) contributions from the Dirac gauginos to the Higgs soft mass
squared(s).  The largest contributions to the soft mass squareds for
the Higgs doublets come from the Dirac wino, 
\begin{equation}
m_{H_u}^2 \simeq m_{H_d}^2 \simeq \frac{g^2}{4\pi^2} M_{D,2}^2 \simeq 
  \left( \frac{M_{D,2}}{10} \right)^2 \, . 
\label{eq:higgssoftsquared}
\end{equation}
Comparing this to the size of $B_\mu$ given in Eq.~(\ref{eq:genBmu}), 
we need $\lambda_{B_\mu} \simeq 10^{-2}$ 
such that $B_\mu$ marginally destabilizes the Higgs mass matrix giving one 
negative eigenvalue.  Since this contribution to $B_\mu$ 
arises in the superpotential, it is technically natural for
this coefficient to be small. 

Notice also that when the one-loop finite contributions from the
Dirac gauginos dominate the Higgs mass squareds, Eq.~(\ref{eq:higgssoftsquared}),
$m_{H_u}^2 \simeq m_{H_d}^2$ and thus $\tan\beta \simeq 1$.  
Once $F$-term contributions are included
with different coefficients for the up-type and down-type masses, 
$\tan\beta$ can be different from $1$.  Generically, in the absence
of large hierarchies in these coefficients, $\tan\beta$ 
is small.

\section{Gauge coupling unification at one-loop}
\label{sec:simpleunify}

We now discuss gauge coupling unification in 
Split Dirac Supersymmetry models.  This discussion
provides a common framework that illustrates the relevant 
contributions to the $\beta$-functions, at one-loop, and the 
expected scales of the superpartners.  In the specific models
described in Secs.~\ref{sec:puredir} and \ref{sec:mbino}, 
we numerically evaluate 
gauge coupling unification to two-loops with the appropriate
thresholds for the spectra in each theory.

Since the sfermions fill out complete GUT multiplets, they do not affect the differential running of the gauge couplings, 
and consequently the level of unification, and we will omit them from the discussion below.  These effects are included in the numerical analysis carried out in later sections.
Thus, there are two important contributions to the one-loop beta-functions
for the gauge couplings that determine the level of unification:  Higgsinos (and Higgses) and gauginos.
Given that $\mu$ is small in Split Dirac Supersymmetry models (Eq.~\ref{eq:mudim7}),
the only difference from MSSM running is the (lack of) gauginos and the 
scalar components of one Higgs doublet.  Since the Higgs scalar doublet 
has a small contribution to the $\beta$-functions, here we focus
on just the gauginos.  

The solutions to the one-loop gauge coupling RGEs in the MSSM are 
\begin{equation}
\alpha^{-1}_{\rm unif}(\Lambda_{\rm unif}) - \alpha^{-1}_i(\Lambda_{\rm weak}) = 
    \frac{b_i}{2\pi}
    \log \left( \frac{\Lambda_{\rm unif}}{\Lambda_{\rm weak}} \right) \, ,
\label{eq:alpha1}
\end{equation} 
where $b_i = b_i^{\rm MSSM} = (33/5, 1, -3)$ are the one-loop beta-function
coefficients of the MSSM, where unification is achieved to within about $1\%$.  
Compare this with Split Dirac Supersymmetry, 
\begin{equation}
\alpha^{-1}_{\rm unif}(\Lambda_{\rm unif}) - \alpha^{-1}_i(\Lambda_{\rm weak}) = 
    \frac{b_i^{\rm Dirac}}{2\pi}
    \log \left( \frac{\Lambda_{\rm unif}}{M_D} \right) + 
    \frac{b_i^{\rm MSSM} - b_i^{\rm gaugino}}{2\pi}
    \log \left( \frac{M_D}{\Lambda_{\rm weak}} \right) \, ,
\label{eq:alpha2}
\end{equation} 
where $b_i^{\rm Dirac} = b_i^{\rm MSSM} + N_i$ and 
$b_i^{\rm gaugino} = 2 N_i/3$, with $N_i = 0,2,3$
the quadratic Casimir for $U(1)_Y$, $SU(2)_L$, $SU(3)_c$. 
The scale for the Higgsinos is assumed to be same ($= \Lambda_{\rm weak}$)
for both Eq.~(\ref{eq:alpha1}) and (\ref{eq:alpha2}).
The additive factor, $b_i^{\rm Dirac}$, corresponds to the 
usual gauginos as well as the fields in the chiral adjoint superfields.
The RGE can be rewritten as
\begin{equation}
\alpha^{-1}_{\rm unif}(\Lambda_{\rm unif}) - \alpha^{-1}_i(\Lambda_{\rm weak}) = 
    \frac{b_i^{\rm MSSM}}{2\pi}
    \log \left( \frac{\Lambda_{\rm unif}}{\Lambda_{\rm weak}} \right) + 
     N_i \frac{1}{2\pi}
    \log \left( \frac{\Lambda_{\rm unif}}{M_D} \right) - 
    \frac{2}{3} N_i \frac{1}{2\pi}
    \log \left( \frac{M_D}{\Lambda_{\rm weak}} \right) \, .\nonumber
\end{equation} 
Crucially, the additive contribution above the scale $M_D$ and
the subtracted contribution below $M_D$ are both proportional to 
the quadratic Casimir of the $i^{\rm th}$ gauge group, $N_i$.
We can solve for the scale $M_D$ where the last two terms cancel 
against each other, 
\begin{equation}
M_D = \Lambda^{3/5}_{\rm unif}\, \Lambda^{2/5}_{\rm weak} \qquad
\mbox{(one-loop)} \, .
\end{equation}
Notice that one obtains the same $M_D$ for all three SM gauge groups -- 
this occurred because the new matter that we added was in the same 
representation as the gauginos.  Setting $\Lambda_{\rm weak} = \tev$, 
which corresponds to a unification scale of $2 \times 10^{16}\, \gev$, 
we find $M_D \sim 10^{11}\, \gev$.

Having determined that the mass scale $M_D$ necessary for Dirac supersymmetry to unify coincides with the scale where the SM Higgs quartic coupling vanishes, and that a vanishing Higgs quartic is a natural boundary condition in Dirac supersymmetry, we are ready to consider specific models. In the following sections we present two complete models within the Split Dirac Supersymmetry framework, each utilizing a subset of the tools presented in Sec.~\ref{sec:toolkit}.  We will find that the two-loop contributions to the gauge coupling evolution cause the unification scale to increase to $\gsim 10^{17}$~GeV and the precision of unification to slightly worsen, which we will quantify.  (And intriguingly, this also occurs in Ref.~\cite{Hall:2013eko}.)

\section{Pure Dirac Model}
\label{sec:puredir}

The first model we consider is one where all the gauginos acquire a
Dirac mass.  We construct the model from 
the relevant toolkit components, then consider the RG evolution in
detail to self-consistently determine the mass 
scales in the model and the level of gauge coupling unification.

The model assumes the dominant supersymmetry breaking contributions 
arise from $D$-terms, leading to the gaugino masses given in
Eq.~(\ref{eq:diracmass}).  In our numerical evaluations, we take
the gauginos to have common mass $M_D$, for simplicity.
The real part of the adjoint scalars, ${\rm Re}(A_a)$ also acquires 
a mass $\sim M_D$.  
The squarks and sleptons of the MSSM receive a (flavor-blind) 
supersoft contribution to their mass, Eq.~(\ref{eq:finitescalar}).
This mass is a threshold effect and is independent of the scale 
at which supersymmetry breaking is mediated.  Scalar masses may 
also receive contributions from $F$-terms, Eq.~(\ref{eq:fterm}), 
which need not be flavor universal.  The relative size of these 
contributions will determine the exact mass of each sfermion but, 
in the absence of cancellations, they are typically at least as 
heavy as the one-loop finite contributions from the Dirac gauginos.
Anomaly mediation will also generate loop-suppressed Majorana
masses for the gauginos, Eq.~(\ref{eq:gauginoratio}), splitting the 
Dirac gauginos into slightly pseudo-Dirac gauginos.  There are also 
anomaly-mediated contributions to the scalar mass squareds 
(both the real and imaginary parts), though these contributions 
are two-loop suppressed relative to the gaugino mass squared.

In this model, there are two contributions to the $\mu$-term.
One arises from the higher dimension operator involving $D$-terms,
Eq.~(\ref{eq:mudim7}), while the second is from the radiative
generation of $\mu$. As discussed in Sec.~\ref{sec:radiativemu},
the radiative generation is further suppressed by the pseudo-Dirac
nature of the gauginos, roughly
\begin{equation}
\mu_{\rm radiative} 
  \sim \sum_{k=1,2} \frac{g_k^2}{16\pi^2}
                    \sin2\beta
                    \frac{M_{M,\,k}}{M_D}\frac{B_\mu}{M_D} 
  \sim \frac{10^{-7} M_D}{\tan{\beta}} \, .
\label{eq:mumin}
\end{equation}

Summarizing the spectrum, the Pure Dirac model contains nearly 
pure Dirac gauginos with mass $M_D$, squarks and sleptons with 
masses $\tilde{m}$ (we assume $\tilde m \le M_D$), Higgs scalars
with masses $\tilde{m}^2 = m_{H_u} m_{H_d}$, 
a $B_\mu$ term with size $B_\mu \simeq m_{H_u} m_{H_d}$, 
and $\tan\beta = \sqrt{m_{H_u}^2/m_{H_d}^2}$ 
(where $m_{H_u}^2,m_{H_d}^2 > 0$). 
The only light states other than the Higgs boson are the Higgsinos, 
with mass $\sim\, \tev$. This is sketched in Fig.~\ref{fig:spectrum}.

\begin{figure}[t]
\centering
\includegraphics[width=0.65\textwidth]{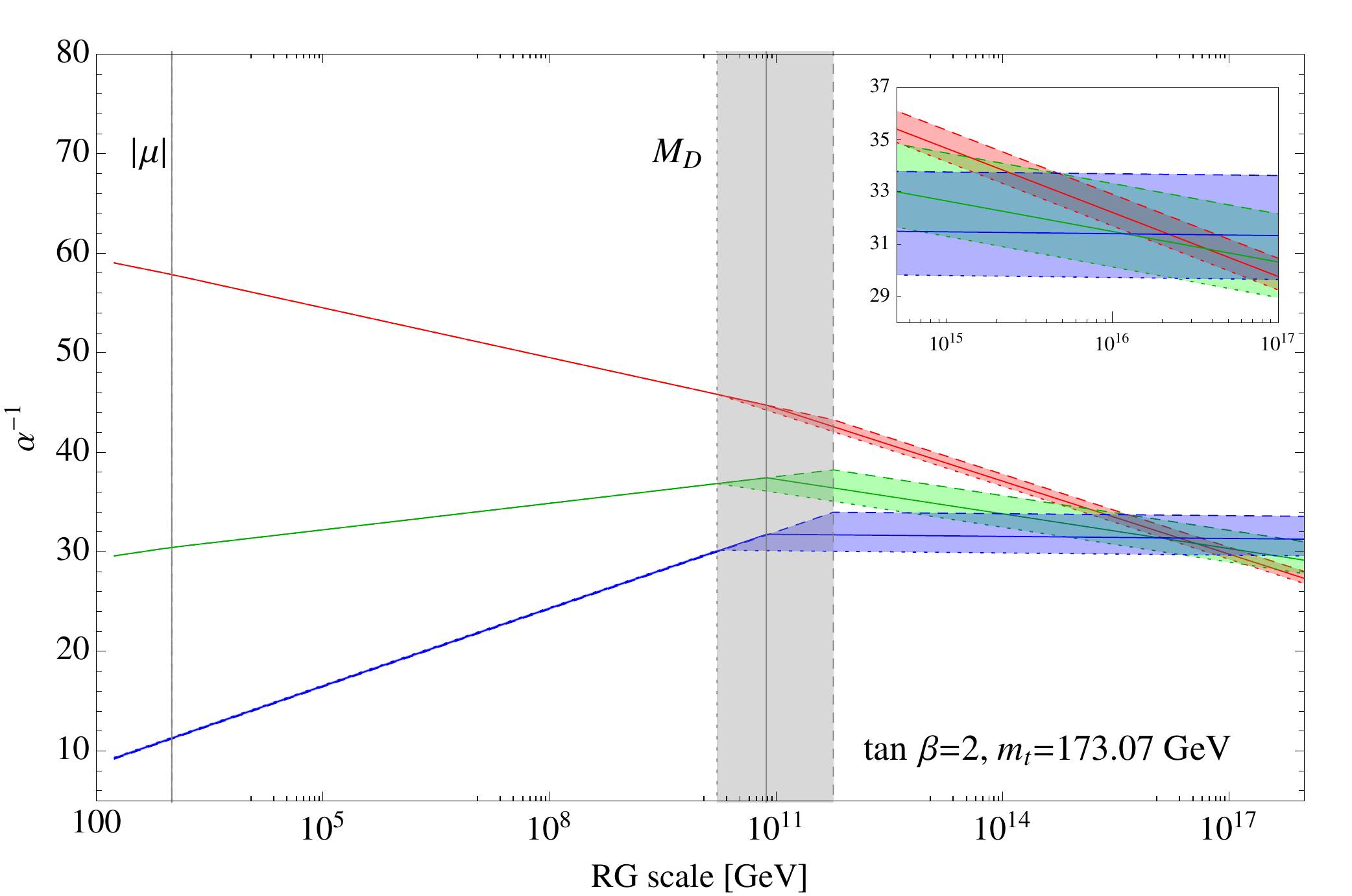}
\caption{An example of running for $\mu\sim1$~TeV and $\tan\beta=2$, the scale 
where the Higgs quartic is zero is $M_D=7.5\times 10^{10}$~GeV\@. The shaded regions correspond to varying $\alpha_s(M_Z)$ within the $2\,\sigma$ uncertainty.}
\label{fig:model1}
\end{figure}

From Sec.~\ref{sec:simpleunify}, we learned that gauge coupling 
unification persists when the Dirac gaugino masses are near the intermediate
scale.  We now carry out a more precise analysis of unification.
In any given model there will be a complicated spectrum with states 
spread from a little above $M_D$ to a loop factor below, with the Higgsinos 
at the TeV scale.  Carrying out the RG evolution in such a scenario 
is a daunting task.  However, the spreading of states over a decade 
or so of energy will not lead to substantially different results from 
the case of degeneracy.  So, for simplicity, we consider a spectrum with 
the Higgsinos and one Higgs light, and all superpartners and the other 
Higgs boson heavy and degenerate, with mass $M_D$.  At the scale $M_D$ 
we match between the non-supersymmetric theory and the MSSM with 
additional adjoints.  We carry out the running of the gauge couplings, 
the top Yukawa, and the Higgs self coupling, at two loops with matching
at tree level.  
(Tree level matching implies the thresholds we discuss are not actually 
physical mass scales but are instead $\overline{\text{MS}}$ masses.)
We follow the approach of 
\cite{Degrassi:2012ry}, which uses results presented in \cite{Broadhurst:1991fy,Melnikov:2000qh,Chetyrkin:1999qi,Hempfling:1994ar,Jegerlehner:2003py}, 
to evolve the couplings from $M_Z$, given in Eq.~(\ref{eq:unificationinput}),
to higher scales using the RG equations applicable to this model, 
given in Appendix~\ref{sec:RGE}.
The scale $M_D$ is determined by the 
renormalization scale where the Higgs self-coupling passes through 
zero\footnote{In this analysis, we assume the contribution from 
Eq.~(\ref{eq:lemontwist2}) is negligible, which is automatic if 
$\tan\beta \simeq 1$.}. 
Under our simplifying assumptions about the spectrum there are very 
few parameters in this model.  Once a Higgsino mass is fixed, there 
is a lower bound on the size of $\tan\beta$ for this Higgsino mass 
to be consistent with the loop generated contribution of Eq.~(\ref{eq:mumin}). 
We show an example of the gauge coupling running in the pure Dirac model 
in Fig.~\ref{fig:model1}.  Note that the level of unification is improved 
as the Higgsino mass is increased.

Because of the large hierarchy between the wino/bino and the Higgsinos 
in this scenario, there is very little mixing among the electroweakinos, 
thus the two (light) neutral Majorana Higgsinos behave essentially 
as a single Dirac fermion.  The relic abundance for a Higgsino in this 
mass range $\sim \tev$, is just right (e.g.~\cite{ArkaniHamed:2006mb}) 
for it to be a thermal DM relic.  Unfortunately, a Dirac fermion that 
has quantum numbers of a neutrino has an unsuppressed elastic scattering 
cross section off nucleons through $Z$ exchange, and is completely 
ruled out by direct detection experiments\footnote{The situation does not improve if the Higgsinos are lighter and do not make up all of the dark matter. The lightest the Higgsinos can be is $\sim 100\,\gev$ (due to the LEP II bound~\cite{lepchargino_lim}), making them only 1\% of the dark matter~\cite{Giudice:2004tc}, while the unsuppressed $Z$-exchange cross section is roughly six orders of magnitude larger than current direct detection limits.}.
So, the Higgsino cannot be the dark matter in this scenario, and therefore must be unstable. 
This can be achieved by either extending to an NMSSM-like scenario where the 
DM is a singlino or by adding $R$-parity violation to make the Higgsino decay, 
with DM coming from another source, \eg\ an axion.  We focus here on the 
latter possibility. 

$R$-parity violating operators fall into two classes, those that violate 
lepton number and those that violate baryon number.  Even with squarks of mass 
$\sim M_D$, there cannot be operators with $\mathcal{O}(1)$ coefficients from 
both classes since this will lead to too rapid proton decay.  

For the single baryon number violating operator, $\lambda_{\slashed B} u^c d^c d^c$,
the Higgsinos will decay via a virtual stop to a top and two jets.  
The partial width for this three-body decay is approximately,
\begin{equation}
\Gamma_{\tilde H} \sim 
  \frac{y_t^2\lambda^2_{\slashed B} \mu^5}{192\,\pi^3 \tilde m^4} \, .
\label{eq:rpvdecay}
\end{equation}
Yielding $\tau_{\tilde H} \sim 7\,$ hours for $\tev$-scale Higgsinos, 
$\lambda_{\slashed B}\sim 1$, and $\tilde m \sim 10^{10}\, \gev$. 
Such long-lived Higgsinos would be completely invisible in collider detectors, 
but there are strong constraints on such long decays from their effects on 
BBN and light element abundances \cite{Kawasaki:2004qu,Kawasaki:2004yh,Jedamzik:2006xz,Jedamzik:2009uy}. This partial width is strongly dependent on mass of squarks and drops $\sim 2$ sec for $\tilde m \sim 10^9\, \gev$.

The results are very similar for the two lepton number violating RPV operators 
$LLe^c$ and $QLd^c$. In the first case the Higgsino decays to $\ell^+\ell^-\nu$ 
and the rate is similar to (\ref{eq:rpvdecay}) suppressed by $(m_\tau/m_t)^2$. 
In the second case the Higgsino decays to a top quark, a down quark and a 
charged lepton and the rate is the same as (\ref{eq:rpvdecay}).

Bilinear R-parity violation may also occur through the lepton number violating 
operator $\kappa_i L_i H_u$.  This can be generated in a similar way to the 
$\mu$-term, of Eq. (\ref{eq:mudim7}), through a K\"{a}hler potential operator of the form 
$\frac{\mathcal W^{'\dag}\mathcal W^{'\dag}}{\Lambda^3} L\, H_u$ 
and so one expects $\kappa\sim\mu\sim 1\,\tev$.
This operator leads to two-body Higgsino decays, 
$\tilde H \rightarrow \ell^\pm W^\mp (\nu Z)$ with a width that scales as,
\begin{equation}
\Gamma_{\tilde H} \sim 
  \frac{g^2}{16\pi} \left( \frac{\kappa\Delta}{\mu^2} \right)^2 \mu \, ,
\end{equation}
where $\Delta$ is the chargino-neutralino mass splitting, which is $\sim 340$~MeV\@.
Usually there are strong constraints on the size of $\kappa_i$ since this 
operator contributes to neutrinos masses at both tree- and 
loop-level \cite{Hall:1983id}.  However, for Dirac gauginos the tree-level
contributions are suppressed by the Majorana mass of the adjoint partner, 
$m_\nu \sim g^2 \langle \tilde{\nu}\rangle^2 M_A/M_D^2$, which we have taken 
to be small.  Furthermore, the loop-generated masses, that arise through the 
mixing of Higgsinos with leptons induced by $\kappa$, scale as,
\begin{equation}
m_\nu \sim \frac{y_b^4}{16\pi^2} \frac{\kappa^2 v_u v_d}{\mu M_D^4} \, .
\end{equation}
Thus, $\kappa\sim 1\,\tev$ is allowed by neutrino masses and leads to 
very fast decays of Higgsinos that are safe cosmologically and can be 
searched for at colliders.

As mentioned above, the $\mu$-term is protected by both a PQ- and an $R$-symmetry, 
so one might worry that turning on RPV interactions leads to a new source 
for generating $\mu$.  The RGEs in a the general MSSM 
with RPV are known up to two-loop order\,\cite{Allanach:1999mh}.  To this order, 
the running of $\mu$ is altered from that of the MSSM only if both 
$\kappa_i$ and one other source of lepton number violation 
(\ie\ $LLE^c$ or $LQD^c$) are non-zero, and the effect is proportional 
to their product.  We ignore these effects.

\section{Hypercharge Impure Model}
\label{sec:mbino}

The Pure Dirac model discussed in the previous section, 
with high scale supersoft supersymmetry breaking, provides an 
explanation of the Higgs quartic coupling crossing through
zero at an intermediate scale (and hence, the correct low energy Higgs mass) 
combined with gauge coupling unification nontrivially obtained through
accelerated running above the intermediate scale.
The downside is that the LSP is not a viable dark matter candidate,
due to the unsuppressed $Z$-exchange with a nearly pure neutral Dirac fermion
made up from the two neutral (Majorana) Higgsinos.  

We now consider a different model, which we dub the Hypercharge Impure model, 
in which
the bino does \emph{not} acquire a Dirac mass, and instead obtains
the standard one-loop suppressed Majorana contribution from 
anomaly-mediation, Eq.~(\ref{eq:amsbgaugino}).  The Majorana bino 
causes a slight splitting 
of the pseudo-Dirac neutral Higgsino into two Majorana 
states.  Consequently, the lightest neutral (Majorana) Higgsino can only 
scatter \emph{inelastically} through $Z$-exchange 
\cite{TuckerSmith:2001hy,TuckerSmith:2004jv,Chang:2008gd},
and thus the spin-independent scattering direct detection rate is suppressed.
If the mass splitting $\gsim 200$~keV, there is negligible scattering through
$Z$-exchange due to insufficient kinetic energy to upscatter into
the heavier neutral Higgsino state.

The absence of a Dirac mass for the bino is automatic if there is
no massless singlet for the bino to marry through Eq.~(\ref{eq:massop})\footnote{If 
the Dirac partners form part of a GUT multiplet, such as a ${\bf 24}$, 
we imagine that the singlet receives a large mass at the scale where the 
GUT breaks and is therefore decoupled from physics at $M_D \ll M_{GUT}$. }.
By itself this does not directly affect gauge coupling unification.
It does, however, have repercussions on the predicted Higgs 
quartic coupling, and consequently, on the mass scales in the model.

In this model, the the wino mass is large ($\sim M_D$), and so 
the neutralino mixing matrix has the form, 
\begin{equation}
\tilde M_N = 
  \left( \begin{array}{ccc} M_1 & -M_Z\,c_{\beta}\,s_W & M_Z\,s_{\beta}\, s_{W} \\ 
                             -M_Z\,c_{\beta}\,s_W & 0 & -\mu \\ 
                              M_Z\,s_{\beta}\, s_{W}  & -\mu & 0 \\ 
         \end{array} \right) \, ,
\end{equation}
with $s_{\beta}= \sin{\beta}, s_W = \sin{\theta_W}$ \etc.  At leading order 
the lightest two (Majorana) eigenvalues are,
\begin{equation}
\tilde M_{N1} = \mu - \frac{M^2_Z\, s^2_W}{2M_1} \, (\sin{2\beta}+1), 
\quad \tilde M_{N2} =  \mu - \frac{M^2_Z\, s^2_W}{2M_1} \, (\sin{2\beta}-1) \, .
\end{equation}
The mass difference is independent of $\mu$ and $\tan{\beta}$ and is
\begin{equation}
\Delta \tilde M_N = \frac{M^2_Z\,\sin^2{\theta_W} }{M_1} 
                  \simeq (200~{\rm keV}) \frac{10^7~{\rm GeV}}{M_1}.
\end{equation}
For spin-independent scattering, and for an inelastic splitting exceeding 
$\gsim 250$~keV, the minimum
velocity to scatter with recoil energy $E_R < 50$~keVnr in xenon is beyond
the maximum velocity any WIMP is expected to have (in the Earth's frame)
assuming a galactic escape velocity of $550$~km/s. 
There is a loop induced spin-independent elastic scattering but again, for these large splittings, 
the rate is much too low to be 
observed~\cite{Hill:2013hoa,Hill:2014yka}\footnote{There is also 
large destructive interference between the $W$-box diagram 
and Higgs exchange at the curiously enigmatic value of 
$m_h \simeq 125$~GeV \cite{Hill:2013hoa}.}.
At tree level, the lightest chargino, the charged component of the Higgsino, 
also has mass $\mu$.  However, there is a loop contribution that splits the 
charged from the neutral component by $\sim 340$ MeV \cite{Cirelli:2005uq}. 
There is also an elastic spin-dependent process, for which the bounds are considerably
weaker, but the rate is suppressed since the coupling scales as 
$\sim \Delta \tilde{M}_N/\mu$.

Compared to the pure Dirac model, the spectrum of the squarks, sleptons 
and Higgs scalars remains relatively unchanged. However, while the pure 
Dirac model was viable even in the limit of zero $F$-term scalar masses, 
the hypercharge impure model is not. If the only source of supersymmetry 
breaking is the supersoft operator, Eq.~(\ref{eq:massop}), removing the 
$U(1)$ adjoint not only leaves both the bino massless, but the the 
right-handed sleptons as well; they are only charged under $U(1)_Y$ 
and would normally receive a mass when the Dirac bino is integrated out. 
The bino mass is lifted from zero by the anomaly-mediated contribution, 
however the anomaly-mediated contribution to the right-handed slepton masses is 
(infamously) tachyonic \cite{Randall:1998uk}. Therefore, there must be 
positive $F$-term contributions to the right-handed slepton masses 
through Eq.~(\ref{eq:fterm}).  To simply the presentation, we assume 
these contributions are comparable to the one-loop finite contributions 
to the other scalars from the Dirac gluino and wino. 

Since the bino mass in this model is purely Majorana, $R$-symmetry 
is broken and $\mu$ will be generated radiatively as soon as supersymmetry 
is broken. The one-loop RG equation for $\mu$ given in Eq.(\ref{eq:genmu2}) 
must be integrated from $B_{\mu}$ all the way down to $M_1$, a much larger 
interval than in the pure Dirac case. 
The larger running interval leads to substantially larger radiative $\mu$. 
Assuming the primordial $|\mu| \ll |M_1|$, we obtain 
\begin{equation}
\mu \; \simeq \; \frac{\tilde{g}'_u \tilde{g}'_d}{16\pi^2} M^*_1 
                 \ln \frac{|B_\mu|^{1/2}}{|M_1|} 
    \; \simeq \; (1 \; {\rm TeV}) \sin{(2\beta)} \, 
                 \frac{M_1^*}{10^6 \; {\rm GeV}} \ln \frac{|B_\mu|^{1/2}}{|M_1|} 
\label{eq:genmu2approx}
\end{equation}
Depending on $M_1$ and $\tan{\beta}$, the generated $\mu$ 
can easily exceed $1\,\tev$.

One additional significant consequence follows from the presence
of a pure Majorana bino.  As shown in Eq.~(\ref{eq:adjointquartic}),
\begin{equation}
\lambda_h(M_D) = \frac{g'^2}{4} \cos^2 {2\beta}
\label{eq:BC}
\end{equation}
and thus a partial quartic coupling is re-generated.  This tends to
lower the scale of the Dirac gauginos (and the other derived scales), 
as we show in more detail in the next subsection.

\subsection{Gauge coupling unification}
\label{sec:whataboutunification}

We now study gauge coupling unification in this model, again using the 
weak scale coupling inputs given in Eq.~(\ref{eq:unificationinput}).  
The RG evolution is done similarly to the Pure Dirac model.
Choosing a Higgsino mass $m_{\tilde{H}} \simeq |\mu|$, 
we evolve the RG equations from the weak scale up to $m_{\tilde H}$, 
and then continue to evolve until the running Higgs quartic coupling 
$\lambda_h$ satisfies the boundary condition\footnote{Like the analysis
for the Pure Dirac model, we assume the contribution from
Eq.~(\ref{eq:lemontwist2}) is negligible.}
\begin{equation}
\lambda_{h, SM+\tilde{H}}(M_D) =
  \frac{g'^2_{SM+\tilde{H}}(M_D)}{4} \cos^2 2\beta \, .
\label{eq:model2bc}
\end{equation}
This sets the Dirac wino mass scale, $M_D$, which we
take to be the same value for the Dirac gluino. 
The subscript in the above equation indicates that the $\lambda_h$ and 
$g'$ RGEs contain the effects of all SM fields plus the Higgsinos. 
This change in the $\lambda_h$ boundary condition is the major difference 
between the RG evolution in this model and the Pure Dirac model 
discussed in Sec.~\ref{sec:puredir}.

Having established $M_D$, we set $M_1 = f M_D$, and we consider $f \in \{ 10^{-4}, 10^{-3}, 10^{-2}, 10^{-1} \}$.  
The range arises from Eq.~(\ref{eq:gauginoratio}), 
where $f \simeq 10^{-2}$ is predicted if $\Lambda = M_{\rm Pl}$,
the couplings $\lambda_{2,3} = 1$ in Eq.~(\ref{eq:massop}), 
and the $D$-term dominates the supersymmetry breaking
contributions in the hidden sector.  Smaller (or larger) values
of $f$ are easily possible, e.g., when $\Lambda < M_{\rm Pl}$ 
(or when $\lambda_{2,3} < 1$).
Generically we expect the squarks and sleptons to be somewhat lighter 
than $M_D$, however for presentation purposes we have 
set $\tilde{m} = M_D$ to minimize the number of thresholds we have to deal with. 
With $M_1$ and $M_D$ (and our assumption about $\tilde m$), 
all thresholds are known, and we can complete the RG evolution
up to and past these mass scales with suitable matching.

\begin{figure}[t]
\centering
\includegraphics[width=0.49\textwidth]{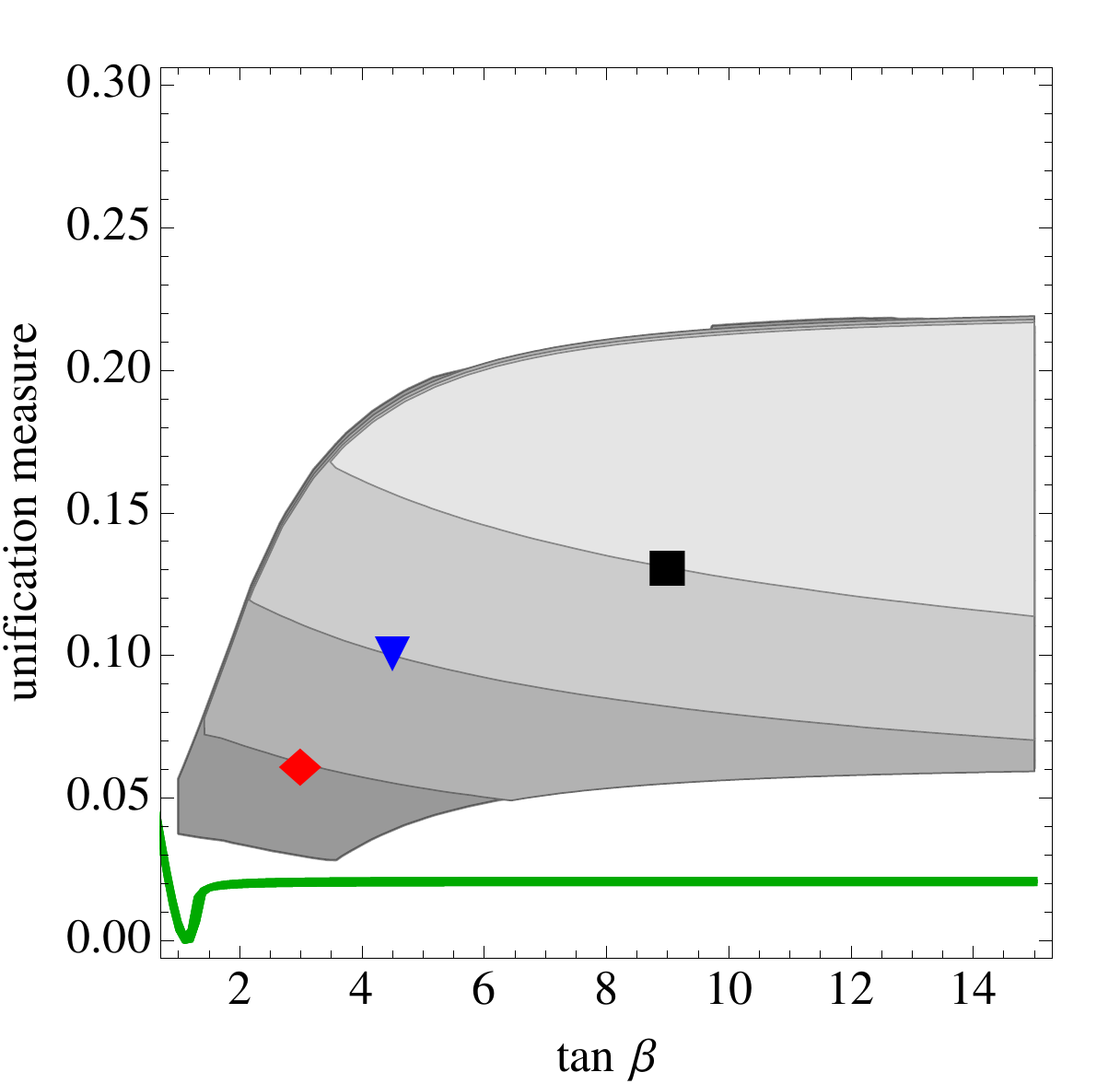}
\includegraphics[width=0.49\textwidth]{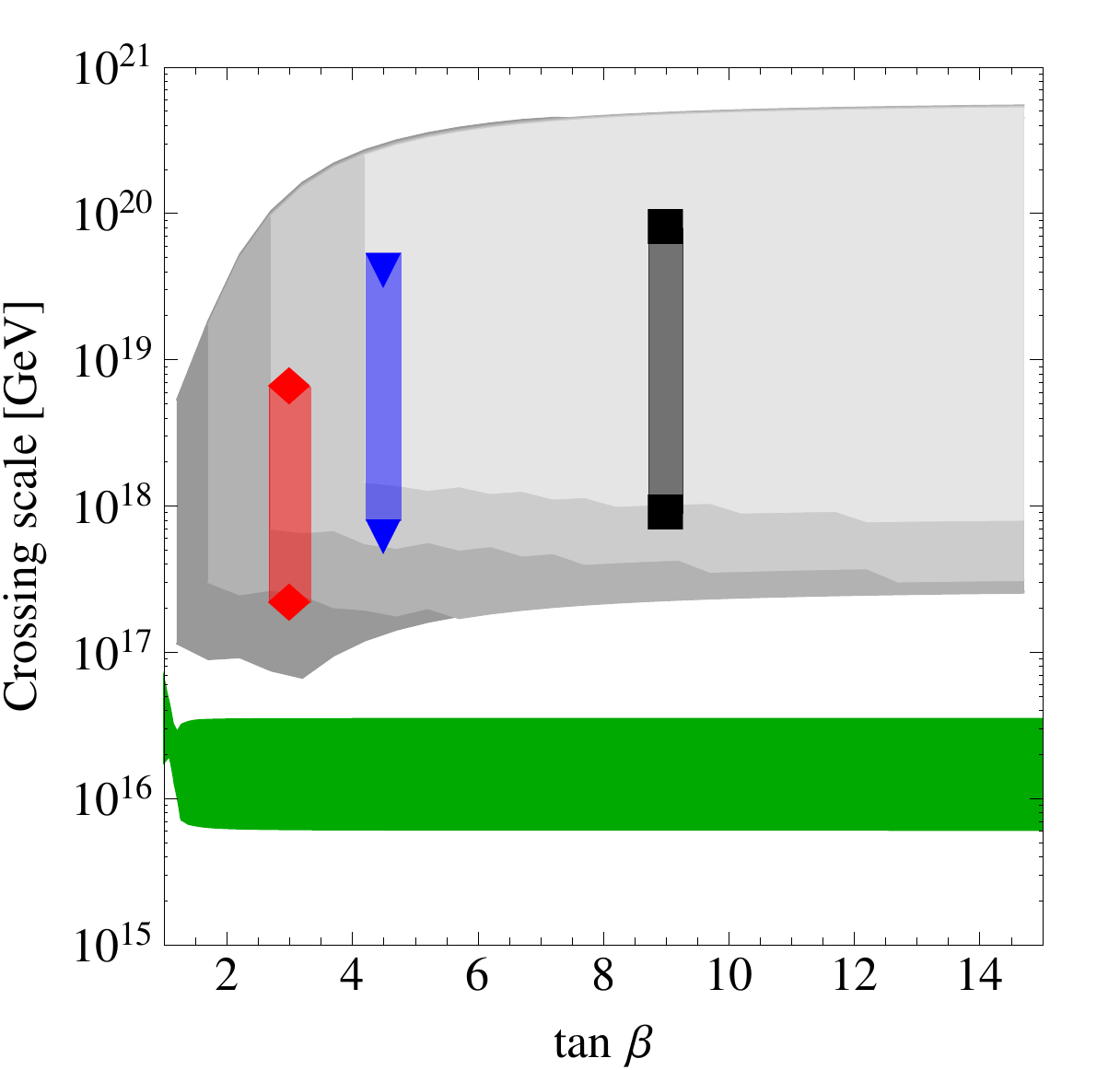}
\caption{The left-side plot shows the unification measure, 
defined as the area of the triangle formed 
by the three gauge coupling intersection points, for four different values of
$M_1/M_D$ as $m_t$ and $\tan(\beta)$ are varied. 
The right-side plot shows the gauge coupling unification scale \emph{range},
defined by the lowest and highest scale where two of the three couplings 
cross each other.  
To scale out the dependence 
of the unification measure on $\alpha^{-1}_{intersect}$, we divide the 
triangle area by the smallest intersection point value of 
$\alpha^{-1}_{intersect}$. Only points with consistent Higgsino mass 
$\mu < 1.1\, \tev$ are included in the plot. The contours, 
reading from upper right to lower left, correspond to 
$M_1/M_D = 0.1, 10^{-2}, 10^{-3}$ and $10^{-4}$. 
The smallest (largest) $m_{t}$  values correspond to the 
lowest (highest) edge of each contour. 
The three markers indicate benchmark $m_t, \tan({\beta})$ points 
that we will examine in more detail. 
To normalize our definition of the unification measure, we show
the unification measure assuming the MSSM with all sparticles at $1$~TeV\@.}
\label{fig:umeasure}
\end{figure}

\begin{figure*}[t]
\centering
\includegraphics[width=0.49\textwidth]{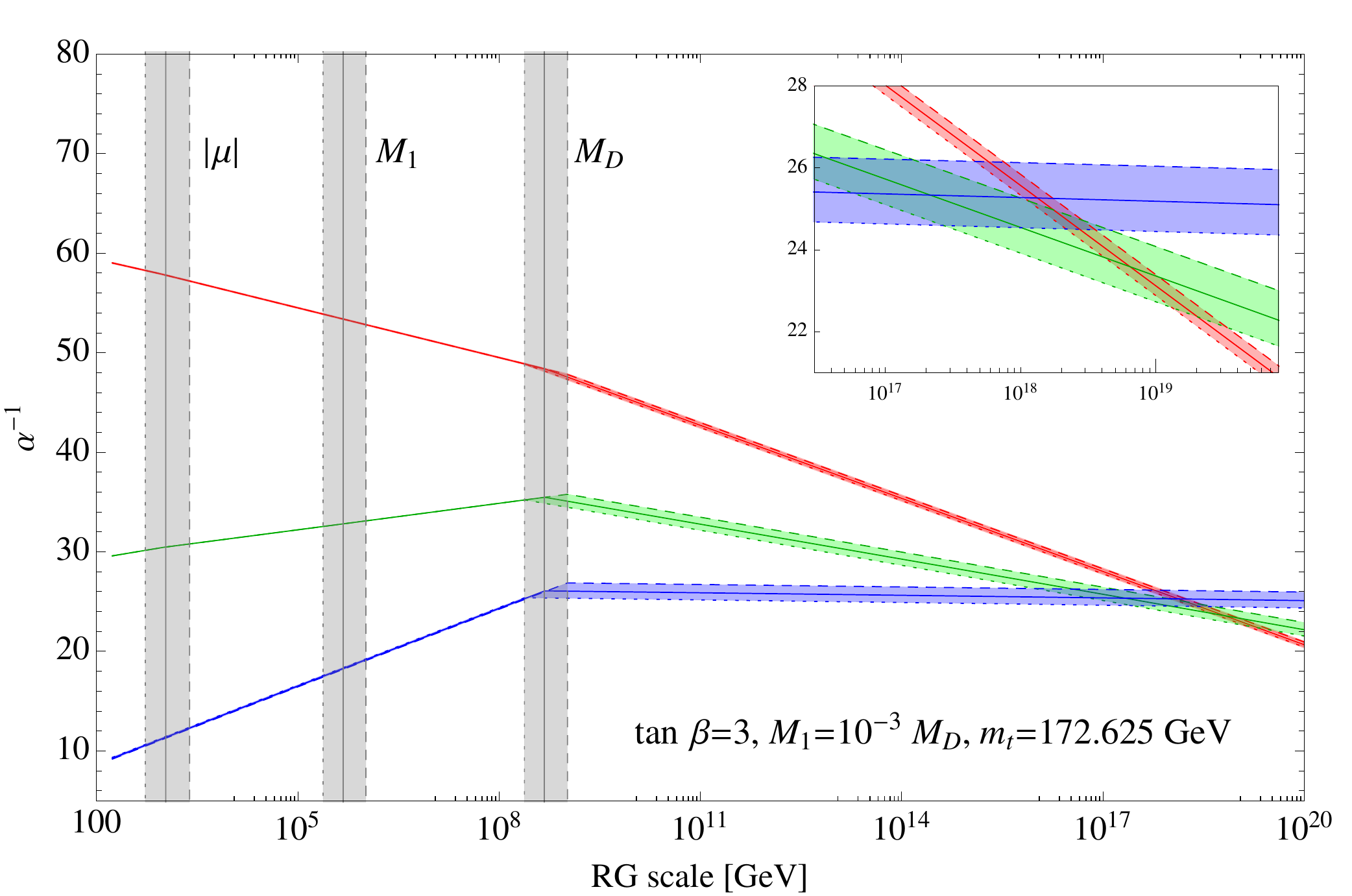}
\includegraphics[width=0.49\textwidth]{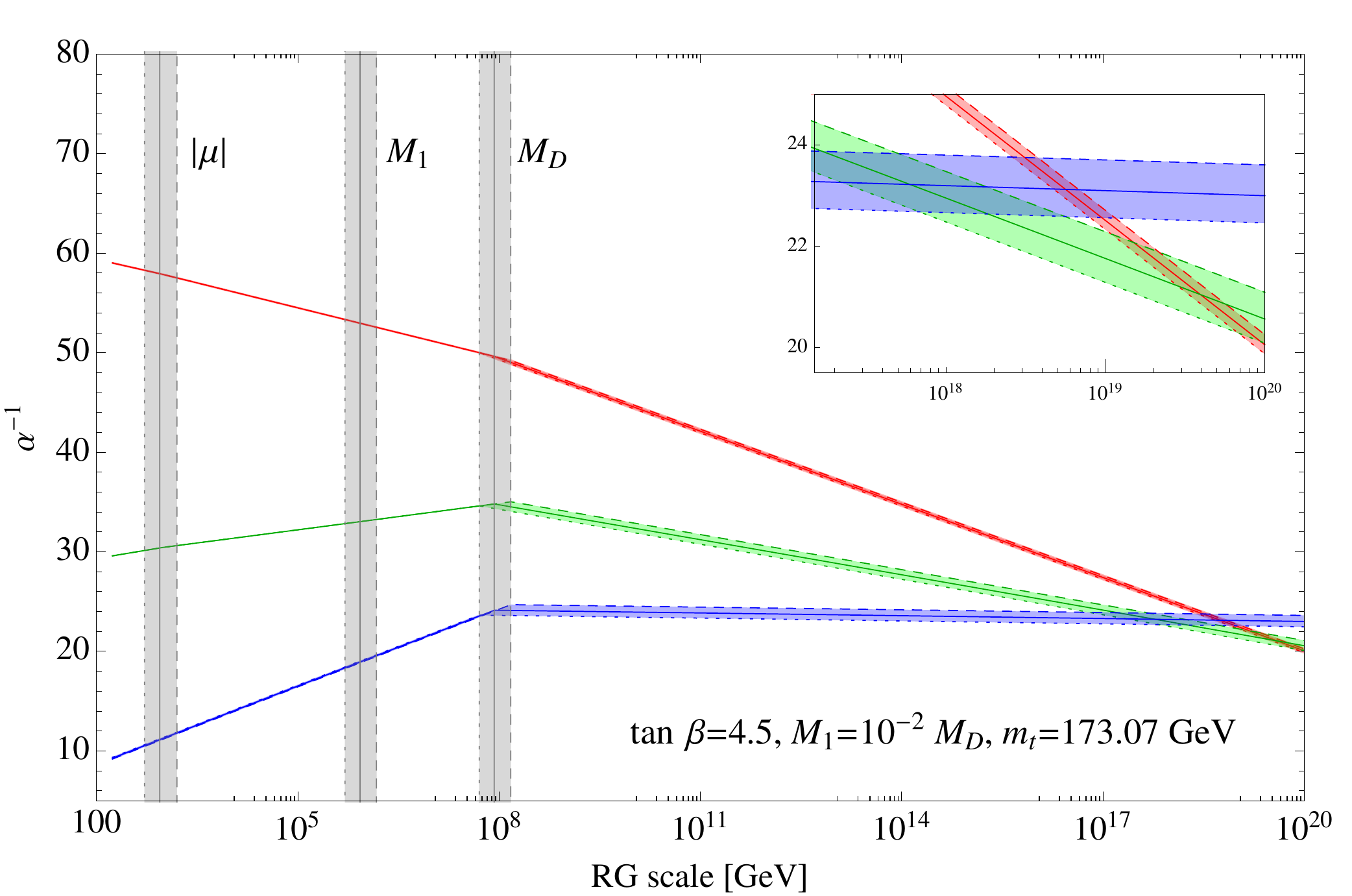} \\
\includegraphics[width=0.49\textwidth]{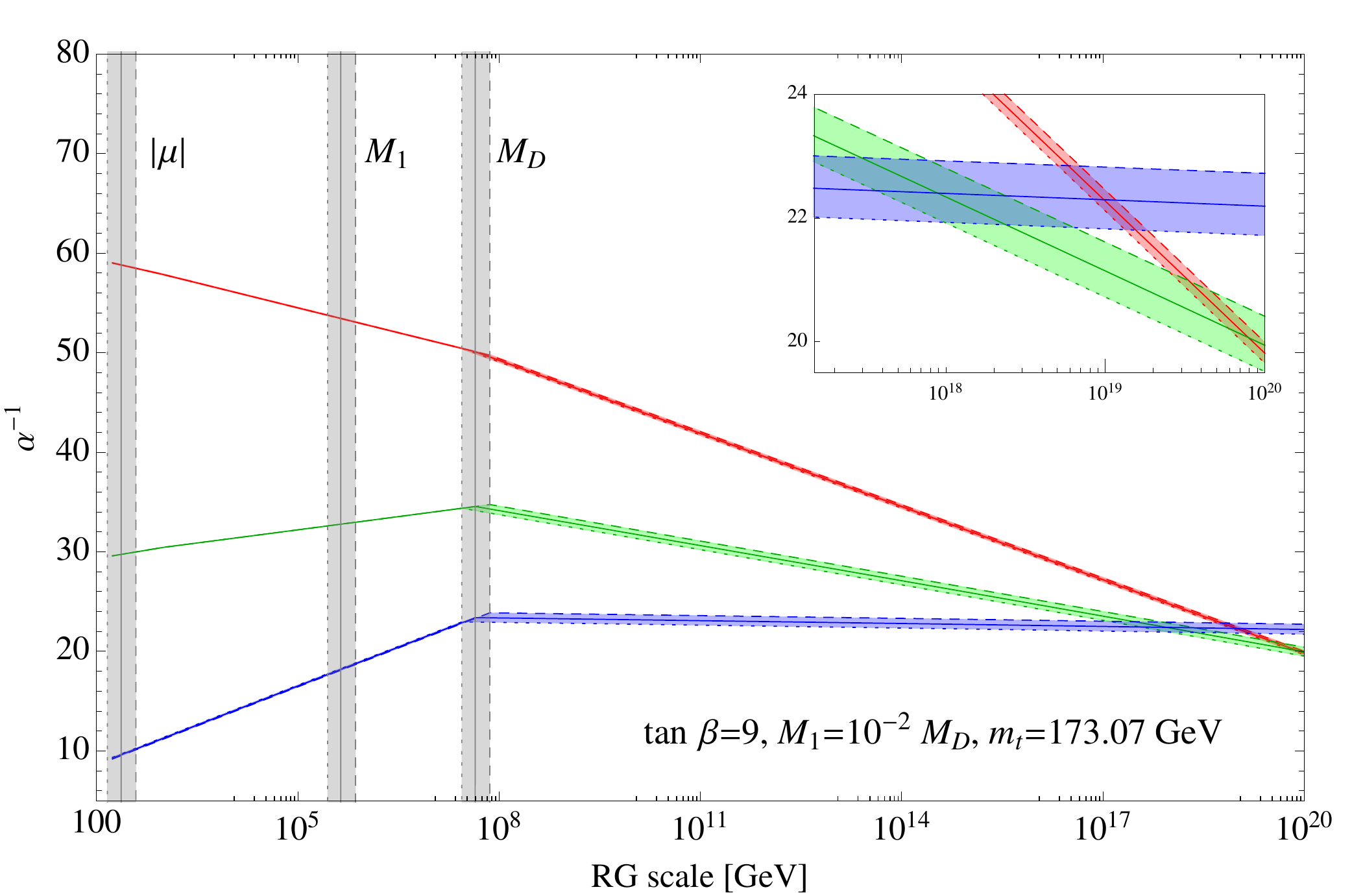}
\caption{The running gauge couplings for the three scenarios indicated by 
markers in Fig.~\ref{fig:umeasure}. The top left plot corresponds to the 
red diamond in Fig.~\ref{fig:umeasure}, the top right plot corresponds to the 
blue triangle, and the lower plot corresponds to the  black square. 
In each scenario we show the variation in the unification as the 
strong coupling $\alpha_s$ is varied within $2\sigma$ of its central value. 
The insets in the upper right of each plot show a zoomed-in picture 
of the intersection region. As explained in the text, since our procedure 
for setting $M_D$ depends on the running of the Higgs quartic, 
all mass scales, and hence all couplings, shift as $\alpha_s(M_Z)$ is varied.}
\label{fig:benchmarkRGE}
\end{figure*}
Finally, to check the consistency of our Higgsino mass choice, we also run 
from UV to IR. Starting at $M_D$ and assuming $\mu(M_D) = 0$, we solve for 
the radiatively generated $\mu$. The choice $\mu(M_D) = 0$ is somewhat arbitrary, 
as we have seen that there can be $O(\tev)$ contributions to $\mu$ from the
higher-dimensional operator shown in Eq.~(\ref{eq:mudim7}). 
A contribution to $\mu$ at the scale $M_D$ is multiplicatively renormalized. 
For the values of $M_1$ that we consider, the effect of the 
multiplicatively renormalized piece of $\mu$ is small, 
however it is possible to arrange for cancellations between this piece 
and the contribution to $\mu$ coming from $M_1$. Some of this possible
parameter space is already incorporated by the large range in $f = M_1/M_D$.

The quantities we are interested in for a given set of inputs are: 
i.)   the ``quality'' of the gauge coupling unification, 
ii.)  the scale of gauge coupling unification, and 
iii.) the internal consistency of the Higgsino mass.

The quality of unification is a somewhat subjective measure;
we choose to calculate the area of the triangle formed, in the usual $\log(\mathrm{RG\ scale})-\alpha^{-1}$ plane, from the 
three coupling intersection points, i.e., where 
$\alpha^{-1}_3 = \alpha^{-1}_1$, $\alpha^{-1}_3 = \alpha^{-1}_2$, etc. 
Each intersection point is a coupling value $\alpha^{-1}_{\rm intersect}$ 
and an energy scale. The area of the triangle is not an ideal measure, 
since it leads to artificially low values for scenarios that happen 
to unify at small $\alpha^{-1}_{\rm intersect}$. Therefore, to remove this bias 
and get a more robust unification measure, we divide the area of the 
unification triangle divided by the smallest of the three 
$\alpha^{-1}_{\rm intersect}$.  To study how the unification scale changes 
through parameter space, we keep track of both the lowest and 
highest energy scales among the intersection points.
Finally, we have also calculated the unification measure and range
of scales in the MSSM, to directly compare with our model.

\begin{figure*}[t]
\centering
\includegraphics[width=0.49\textwidth]{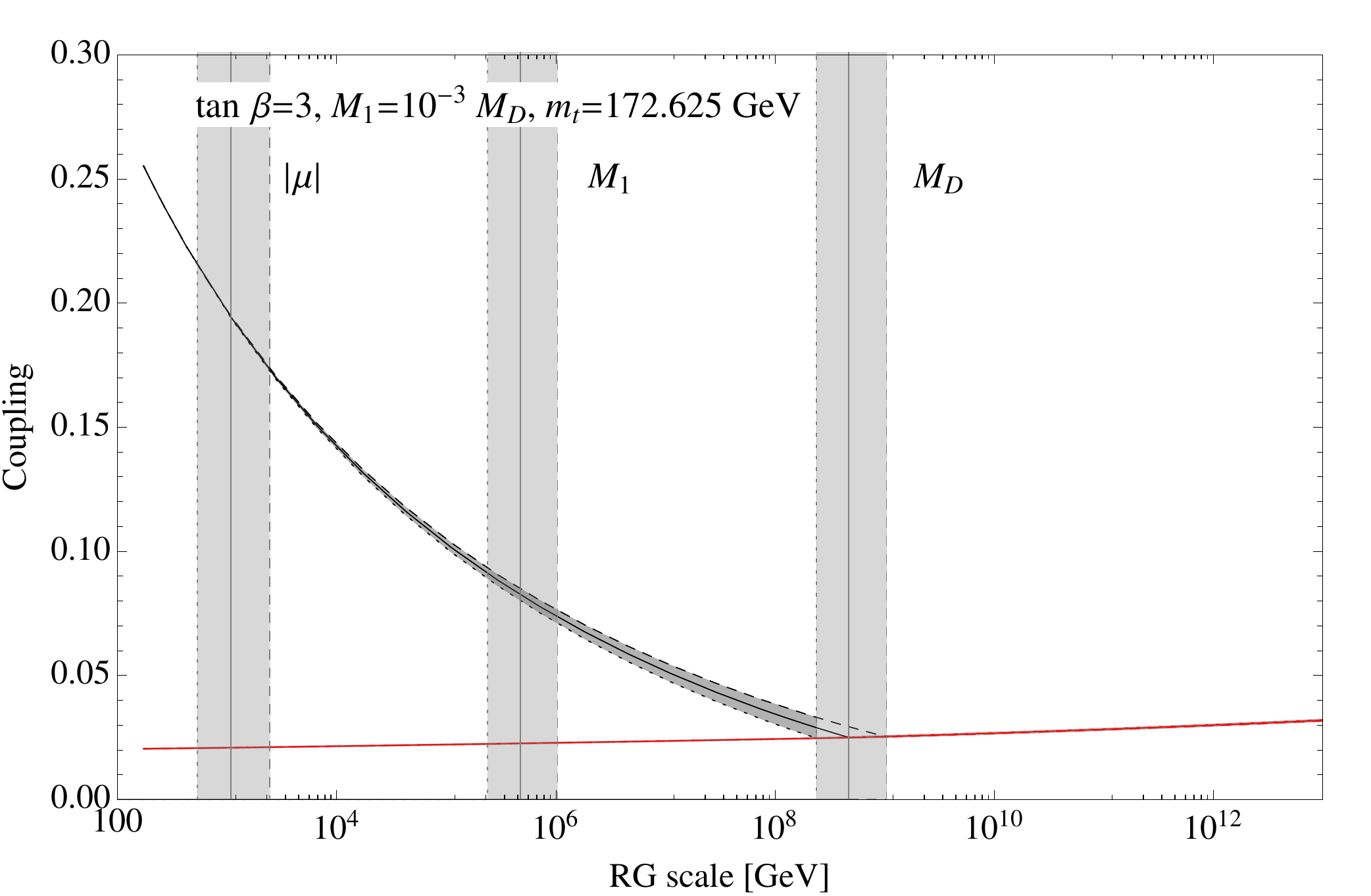}
\includegraphics[width=0.49\textwidth]{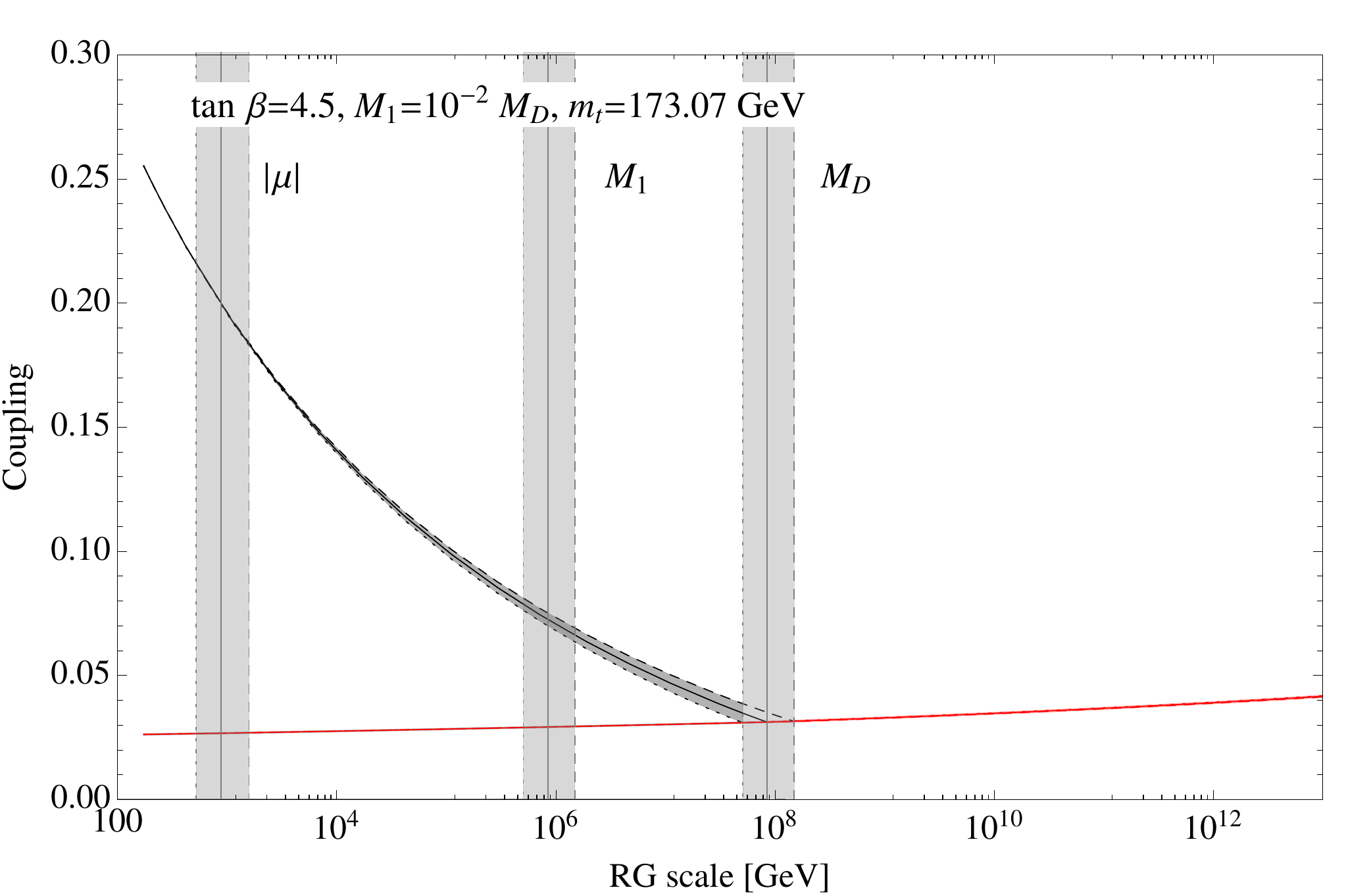}\\
\includegraphics[width=0.49\textwidth]{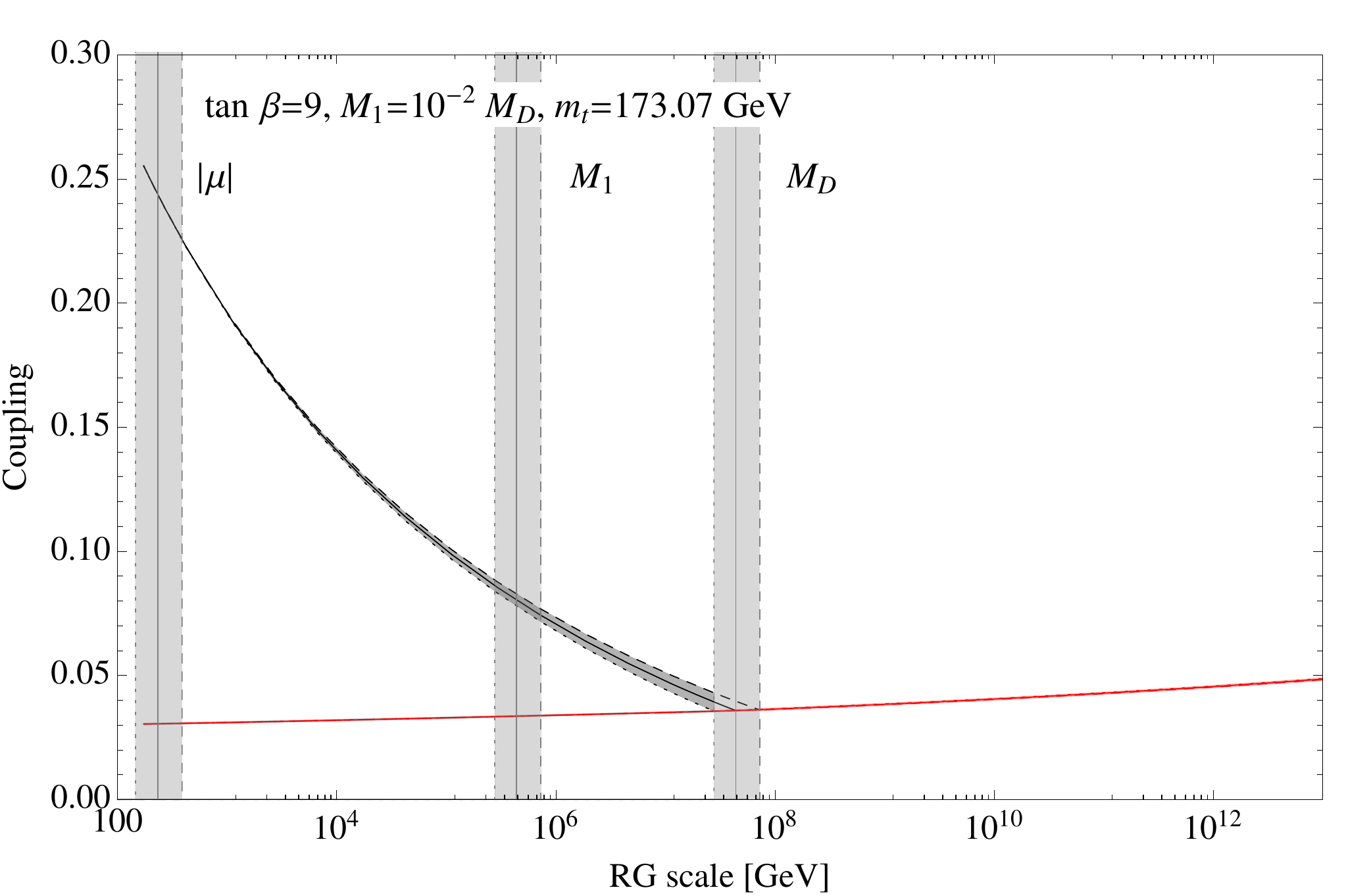}
\caption{The running of the Higgs quartic coupling (black) and
$\frac{g'^2}{4}\cos^2 2\beta$ (red) in the three benchmark scenarios 
indicated on Fig.~\ref{fig:umeasure}. The layout of scenarios is the same 
as in Fig.~\ref{fig:benchmarkRGE}.}
\label{fig:benchmarklambda}
\end{figure*}
 
The SM input with the greatest impact on the RG evolution and 
gauge coupling unification is the top mass $m_t(m_t)$. 
A smaller top Yukawa coupling causes the Higgs quartic coupling 
to evolve more slowly, which in turn postpones the 
scale where the quartic and gauge couplings intersect, Eq.~(\ref{eq:model2bc}).
Conversely, a larger top Yukawa coupling causes the Higgs quartic coupling 
to evolve faster and tends to lower the mass scales
in the theory.  We have already seen from the one-loop estimates
Sec.~\ref{sec:simpleunify}, as well as the two-loop results shown in
Fig.~\ref{fig:model1} that gauge coupling unification with a Dirac gluino and wino
prefers $M_D \sim 10^{11}\,\gev$.  Lower $M_D$ 
(due to large $m_t$ values or other effects) causes the gauge coupling
unification to be less precise.  We account for this dependence
by varying $m_t(m_t)$ within the $2\,\sigma$ uncertainty bands in 
our calculations.  The regions formed by varying $m_t(m_t)$ and 
$\tan{\beta}$ are shown in Fig.~\ref{fig:umeasure}. 
As we vary  $m_t(m_t)$ and $\tan{\beta}$ we calculate the (one-loop) radiatively generated $\mu$-term, assuming 
the primordial $\mu$ is 0, and keep only those points for which $\mu\le1.1$ TeV.  The unification measure 
and scale in the MSSM (all superpartners at $1\,\tev$) is also shown 
in Fig.~\ref{fig:umeasure} for comparison.

To give the reader a more concrete context on the quality of gauge coupling 
unification, we pick three benchmark scenarios to display in more detail. 
These three benchmark points are indicated by the markers on 
Fig.~\ref{fig:umeasure}. From the $m_{t}, \tan{\beta}$ and $M_1/M_D$ 
inputs corresponding to each point, we show how the gauge couplings evolve 
with energy, i.e., the analogous plot to Fig.~\ref{fig:model1}. 
The running couplings for the benchmark points are shown 
in Fig.~\ref{fig:benchmarkRGE}. 

As in Fig.~\ref{fig:model1}, we plot the couplings for three different 
choices of $\alpha_s(M_Z)$. The impact of varying $\alpha_s(M_Z)$ 
is larger than one might have expected; all couplings and scales move, 
some even significantly, as $\alpha_s(M_Z)$ is varied. 
This sensitivity comes from the fact that we use the running Higgs quartic, 
a quantity sensitive to $\alpha_s(M_Z)$, to set the location of $M_D$. 
Small changes in $\alpha_s(M_Z)$ can lead to $O(1)$ changes in what we 
derive $M_D$ to be, and changes in $M_D$ trickle down to changes 
in where all running couplings are matched. To better illustrate how 
the scale $M_D$ is derived, and how changes in $\alpha_s(M_Z)$ affect it, 
we plot the running quartic coupling in each of the 
benchmark scenarios in Fig.~\ref{fig:benchmarklambda} below. 
Along with $\lambda_h$, we also show the running of 
$\frac{g'^2}{4}\cos^2 2\beta$, as the intersection of the two curves 
is what sets $M_D$.

We can see from Fig.~\ref{fig:umeasure} that, at low $\tan{\beta}$ and small $M_1/M_D$, unification can be as good as in the MSSM. For other parameters, unification is somewhat less precise. The inset plots in Fig.~\ref{fig:benchmarkRGE} show the mismatch in unification after uncertainties in $\alpha_s(M_Z)$ are taken into account.

\section{Discussion}
\label{sec:discussion}

We have presented a new framework for split supersymmetry employing
Dirac gaugino masses at intermediate mass scales ($\sim 10^{8-11}\,\gev$).  
Two specific models were constructed, both containing gauge coupling 
unification, and one (the Hypercharge Impure model) 
with a Higgsino dark matter candidate.  There are no
model-building gymnastics necessary to suppress $R$-violation to maintain 
light gauginos, as in the original split supersymmetry model. 
The predictivity of Split Dirac Supersymmetry is improved
over Simply Unnatural / Mini-Split / Spread, see for example Refs. \cite{Hall:2011jd,Ibe:2012hu,Arvanitaki:2012ps,Hall:2012zp,ArkaniHamed:2012gw}, 
in so far as the split superpartner mass scale is determined to be
an intermediate scale with a weaker dependence on $\tan\beta$.  

Both of the discussed models have signals at the weak scale.  We emphasize
that the signals themselves are qualitatively distinct from other split supersymmetry models -- 
just Higgsinos are light in Split Dirac Supersymmetry, while
binos, winos, gluinos are \emph{heavy}. 
One of the pressing issues of models that implement the scalar-to-gaugino
mass hierarchy using anomaly mediation \cite{Giudice:1998xp,Wells:2004di},
that can also occur in the Refs.~\cite{Hall:2011jd,Ibe:2012hu,Arvanitaki:2012ps,Hall:2012zp,ArkaniHamed:2012gw}
and related models \cite{Cheung:2005ba}
is that wino-like dark matter is strongly constrained by indirect detection
from $\gamma$-ray production in the center of the galaxy 
\cite{Cohen:2013ama,Fan:2013faa}.
Nearly pure Higgsino-like dark matter, with a mass of $\simeq 1.1$ TeV (which is consistent with thermal abundance), does not suffer from this 
constraint due to the negligible Sommerfeld enhancement in the 
annihilation rate.  On the contrary, indirect detection may provide one of the promising 
avenues towards experimental verification \cite{FKMpreparation}.
There are several other aspects of split supersymmetry,
including flavor physics \cite{Altmannshofer:2013lfa,Baumgart:2014jya}
and inflation \cite{Craig:2014rta_beach_and_sunshine_are_overrated} 
that could have interesting
interpretations in the Split Dirac Supersymmetry framework. 

Another challenge to split supersymmetry models is dimension-5 proton decay 
with anarchic sfermion masses \cite{Dine:2013nga,Nagata:2013sba}.  
Split Dirac Supersymmetry with just $D$-term supersymmetry breaking mediation 
is flavor-blind, completely eliminating this issue.  Nevertheless, even if 
anarchic $F$-terms are also present (and $F$-terms must be present in the 
Hypercharge Impure model), the situation with Split Dirac Supersymmetry
is much improved because of the absence of Majorana gluinos and winos, 
and that the only $R$-violation arises from a one-loop suppressed
bino mass that is accompanied by its small $g'$ couplings to sfermions.

Gauge coupling unification is comparable to the MSSM in the Pure Dirac model,
and somewhat worse in the Hypercharge Impure model.  Since the predicted
unification scale is higher $\gsim 10^{17}$~GeV, the Planck-suppressed
GUT threshold corrections are also correspondingly larger.  
Hence, the slightly less precise unification could be just 
a symptom of this higher GUT scale.  Among the three scales where the gauge
couplings intersect, $\alpha_2 = \alpha_3$ occurs at the lowest scale
with $\alpha_1$ typically $\sim 5$\% smaller at this scale. 
If we take this minor discrepancy as suggestive of low energy physics, 
this could suggest the sleptons are actually much lighter than the 
squarks in the model.  Such a spectrum is not unexpected since in the Hypercharge Impure model
the source of the (RH) slepton mass is distinct from that for the squarks.
There may also be additional fields transforming 
under $U(1)_Y$ at low to intermediate scales.

One of the most striking results from our study is the possibility
of nearly pure weak scale Higgsino dark matter whose mass $\mu$ and 
neutral Higgsino splitting $\Delta \tilde{M}_N$ arise from the same 
source -- a large Majorana bino mass $M_1 = 10^6 \ra 10^7$~GeV\@.  
Split Dirac Supersymmetry (in the Hypercharge Impure variety) acts 
as a UV completion of \emph{viable} Higgsino dark matter.  
Higgsino dark matter produced purely from thermal processes in the
early Universe is possible when $\mu \simeq 1.1$~TeV, though 
lighter Higgsinos are also possible if there is an additional 
source, e.g., asymmetric Higgsinos \cite{Blum:2012nf} 
or a non-thermal source \cite{Allahverdi:2012wb}.

There are numerous phenomenological consequences of Higgsino
dark matter that warrant a separate study, which we will present
in Ref.~\cite{FKMpreparation}.  
On the dark matter side, we would like to know how best to 
detect an inelastically split Higgsino.  Direct detection
is highly suppressed, however, there can be a large degree of time  
and recoil-energy dependence that, to the best of our knowledge, 
are not being searched for now with existing data. 
Direct detection through elastic scattering is  
highly suppressed for both the loop induced processes leading 
to spin-independent scattering as well as the tree level spin-dependent
scattering (due to pseudo-Dirac nature of the lightest Higgsino).
Indirect detection through $\gamma$-rays provides a promising 
detection strategy using proposed future air Cherenkov telescopes
\cite{CTA}. Indirect detection through accumulation and annihilation
in the Sun \cite{Menon:2009qj,Nussinov:2009ft,Shu:2010ta}, 
white dwarf \cite{McCullough:2010ai} also provide interesting probes.
However, thermalization of dark matter has been \emph{assumed} in
Refs.~\cite{Menon:2009qj,Shu:2010ta,McCullough:2010ai}
and unfortunately, the highly suppressed spin-independent 
elastic scattering suggests thermalization is not effective on 
timescales of order the age of the solar system. 
On the collider side, pure Higgsinos are currently unconstrained
by the LHC  \cite{Han:2014kaa}, beyond the LEP II bound \cite{lepchargino_lim}.  Some first studies of Higgsino production
at the LHC and at a $100$~TeV collider \cite{Low:2014cba} 
suggest getting to the thermal 
abundance upper bound of $1.1$~TeV is not trivial.  Further studies of nearly degenerate Higgsinos are clearly warranted.

\subsection*{Acknowledgments}

We thank N.~Arkani-Hamed, C.~Burgess, P.~Saraswat, N.~Weiner, and I.~Yavin 
for several useful discussions during the course of the research. 
GDK thanks L.~Hall and Y.~Nomura for fun discussions about 
Ref.~\cite{Hall:2013eko} prior to that (and this) paper appearing on arXiv.org.  
GDK is supported in part by the US Department of Energy under 
contract numbers DE-FG02-96ER40969 and DE-SC$0011640$. Fermilab is operated by Fermi Research Alliance,
LLC under Contract No. DE-AC02-07CH11359 with the United States Department of Energy.

\appendix

\section{RG Inputs}
\label{sec:inputs}

For the RG evolution, we take as boundary conditions \cite{PDG},
\begin{eqnarray}
\alpha^{-1}_{em}(M_Z) &=& 127.944 \pm 0.014~,\nonumber \\
\sin^2{\theta_W}(M_Z) &=& 0.23126 \pm0.00005~,\nonumber \\
\alpha_3(M_Z)         &=& 0.1185\pm0.0006~, \\
M_Z(M_Z)              &=& 91.1876\pm 0.0021\, \text{GeV}~, \nonumber\\
m_t(m_t)              &=& 173.07 \pm 0.89\, \gev~, \nonumber\\
m_h                   &=& 125.9\pm0.4\,\gev \, . \nonumber
\label{eq:unificationinput}
\end{eqnarray}

\section{RGE in Dirac-Split}
\label{sec:RGE}

Here we collect the 2 loop renormalisation group equations\footnote{As a check of our method we have derived the 2 loop RGEs for split supersymmetry and agree with the results presented in \cite{Binger:2004nn}.} used to evolve couplings from the top mass to the GUT scale, derived using the standard techniques \cite{Machacek:1983tz,Machacek:1983fi,Machacek:1984zw,Martin:1993zk,Luo:2002ti,Goodsell:2012fm}.  Below the scale $M_D$, the superpartner mass scale, we consider the evolution of the Higgs quartic coupling $\lambda_h$, the top Yukawa $y_t$, and the three gauge couplings $g_i$.  Above that scale we only evolve the gauge couplings and the top Yukawa.  We work in a GUT normalization, $g_1=\sqrt{5/3}\, g_Y$. It is useful to introduce a general form for the two-loop gauge coupling RGEs,
\begin{eqnarray}
\frac{d}{d t} g_i &=& \beta^{(1)}_{i} + \beta^{(2)}_{i} \label{eq:2loopRGE} \\
\kappa\,   \beta^{(1)}_{i}  &=& b_i g_i^3 \nonumber \\
\kappa^2\, \beta^{(2)}_{i}  &=& g^3_i 
    \Big[ \sum\limits_{j = 1}^{3} B_{ij}\, g^2_j - d_i\, y^2_t \Big]~, \nonumber 
\end{eqnarray}
where we define the loop factor $\kappa=16\pi^2$. 
In addition, we define the beta functions 
\begin{eqnarray}
\frac{d}{d t} y_t &=& \beta^{(1)}_{y_t} + \beta^{(2)}_{y_t} \\
\frac{d}{d t} \lambda_h &=& \beta^{(1)}_{\lambda_h} + \beta^{(2)}_{\lambda_h} 
\end{eqnarray}
with coefficients as given below.

\subsection*{Standard Model}

Above the top quark mass, but below the Higgsino mass, the field content is identical to the SM.  Thus,
\begin{equation}
b = \Big( \frac{41}{10}, -\frac {19} 6, -7 \Big),\, \quad B = \left( \begin{array}{ccc} \frac {199}{50} & \frac{27}{10} & \frac{44}{5} \\ \frac{9}{10} & \frac{35}{6} & 12 \\ \frac{11}{10} & \frac 9 2 & -26 \\ \end{array}\right),\, \quad d = \Big(\frac{17}{10}, \frac 3 2,\ 2 \Big)~. 
\end{equation}
Similarly the running of the Yukawa and quartic are as in the SM,
\begin{eqnarray}
\kappa\, \beta^{(1)}_{y_t} &=& \frac 9 2 y^3_t - y_t\,\Big( 8\,g^2_3 + \frac 9 4\,g^2_2 + \frac{17}{20}\,g^2_1 \Big) \\
\kappa^2\, \beta^{(2)}_{y_t} &=& -12\, y^5_t+ y^3_t \Big( 36\,g^2_3 + \frac{225}{16}\, g^2_2 + \frac{393}{80}\,g^2_1 - 6\, \lambda_h\Big)  \nonumber \\
& &{} + y_t\,\Big( -108\, g^4_3 + 9\, g^2_2\,g^2_3 + \frac{19}{15}\, g^2_3\, g^2_1- \frac{23}{4} g^4_2 - \frac{9}{20} g^2_2\, g^2_1 + \frac{1187}{600}\, g^4_1 + \frac 3 2\, \lambda_h^2 \Big) \, ,
\end{eqnarray}
and
\begin{eqnarray}
\kappa\, \beta^{(1)}_{\lambda_h} &=& 12\, \lambda_h^2 + \lambda_h\,\Big(12\, y^2_t - 9\, g^2_2 - \frac 9 5\, g^2_1 \Big) - 12\, y^4_t + \frac 9 4 g^4_2 + \frac{9}{10} g^2_2 g^2_1 + \frac{27}{100}\, g^4_1 \\
\kappa^2\, \beta^{(2)}_{\lambda_h} &=& -78\, \lambda_h^3 + \lambda_h^2\, \Big( 54\, g^2_2 + \frac{54}{5}\, g^2_1 - 72\, y^2_t \Big) + \lambda_h\, \bigg( -3 y^4_t + y_t^2(80\, g^2_3 + \frac {45}{2} g^2_2 + \frac{17}{2} g^2_1) \nonumber \\
& &{} - \frac{73}{8}\,g^4_2 + \frac{117}{20}\,g^2_2\,g^2_1 + \frac{1887}{200} g^4_1 \bigg) + 60\, y^6_t - y^4_t\, \Big(64\, g^2_3 + \frac{16}{5}\,g^2_1 \Big) 
\nonumber \\
& &{} + y^2_t\, \Big( - \frac{9}{2} g^4_2 + \frac{63}{5} g^2_2 g^2_1 - \frac{171}{50} g^4_1 \Big) + \frac{305}{8} g^6_2 - \frac{289}{40} g^4_2 g^2_1 - \frac{1677}{200} g^2_2 g^4_1 - \frac{3411}{1000} g^6_1 \, . \qquad
\end{eqnarray}

\subsection*{The Standard Model with Higgsinos}

The inclusion of the vector-like Higgsinos alters the running of the gauge couplings at one loop, and the quartic and top Yukawa at two loops.  Thus,
\begin{eqnarray}
b = \Big( \frac 9 2, -\frac {5} 2, -7 \Big)\, \quad B = \left( \begin{array}{ccc} \frac {104}{25} & \frac{18}{5} & \frac{44}{5} \\ \frac 6 5 & 14 & 12 \\ \frac{11}{10} & \frac 9 2 & -26 \\ \end{array}\right)\, \quad d = \Big(\frac{17}{10}, \frac 3 2,\ 2 \Big)~.
\end{eqnarray}
Since the one-loop running is as in the SM we only show the two-loop contributions.  First for the top Yukawa,
\begin{eqnarray}
\kappa^2\, \beta^{(2)}_{y_t} &=& -12\, y^5_t+ y^3_t \Big( 36\,g^2_3 + \frac{225}{16}\, g^2_2 + \frac{393}{80}\,g^2_1 - 6\, \lambda_h \Big) \nonumber \\
& &{} + y_t\,\Big( -108\, g^4_3 + 9\, g^2_2\,g^2_3 + \frac{19}{15}\, g^2_3\, g^2_1- \frac{21}{4} g^4_2 - \frac{9}{20} g^2_2\, g^2_1 + \frac{1303}{600}\, g^4_1 + \frac 3 2\, \lambda_h^2 \Big) \, .
\end{eqnarray}
Then the Higgs quartic coupling,
\begin{eqnarray}
\kappa^2\, \beta^{(2)}_{\lambda_h} &=& -78\, \lambda_h^3 + \lambda_h^2\, \Big( 54\, g^2_2 + \frac{54}{5}\, g^2_1 - 72\, y^2_t \Big) + \lambda_h\, \Big(-3 y^4_t + y_t^2(80\, g^2_3 + \frac {45}{2} g^2_2 + \frac{17}{2} g^2_1) \nonumber \\
& &{} - \frac{33}{8}\,g^4_2 + \frac{117}{20}\,g^2_2\,g^2_1 + \frac{2007}{200} g^4_1 \Big) + 60\, y^6_t - y^4_t\, \Big(64\, g^2_3 + \frac{16}{5}\,g^2_1 \Big) + y^2_t\,\Big(-\frac 9 2 g^4_2 \nonumber \\
& &{} + \frac{63}{5}\, g^2_2\, g^2_1 - \frac{171}{50}\,g^4_1 \Big) + \frac{273}{8} g^6_2 - \frac{321}{40} g^4_2\, g^2_1 -\frac{1773}{200}g^2_2\,g^4_1 - \frac{3699}{1000} g^6_1 \, . 
\end{eqnarray}

\subsection*{The Standard Model with Higgsinos and a Bino}

In the second version of the model the bino does not have a adjoint partner to marry and is considerably lighter than the other superpartners.  While the addition of a pure gauge singlet does not alter the running of the gauge couplings directly, the presence of both the Higgsinos and bino as propagating degrees of freedom means there are additional Yukawa couplings we have to consider,
\begin{equation}
\mathcal L \supset \frac{\tilde g'_u}{\sqrt 2} H^{\dag}\, \tilde B \tilde H_u  + \frac{\tilde g'_d}{\sqrt 2} (H^{T}\epsilon) \tilde B \tilde H_d + h.c. 
\end{equation}
These interactions are the supersymmetrization of the $U(1)_Y$ gauge-matter interactions. Had both Higgses been as light as the bino, this piece would have been combined into the supersymmetric $O(g^5)$ piece of the RGE. However, since only one Higgs is (tuned to be) light, the Higgsino-Higgs-bino interactions is instead projected onto that light combination, matched at the bino mass, then run as Yukawa couplings. Matching at this scale $\tilde g'_u = g' \sin{\beta}, \tilde g'_d = g' \cos{\beta}$. These additional Yukawa interactions alter the two-loop gauge RGE, (\ref{eq:2loopRGE}) is modified to become, 
\begin{equation}
\kappa^2\, \frac d {dt}\, g_i = \kappa\,b_i\, g^3_i  + \frac{g^3_i}{(4\pi)^2} \Big[ \sum\limits_{j = 1}^{3} B_{ij}\, g^2_j - d_i\, y^2_t - d_{B,i} (\tilde g'^2_u +  \tilde g'^2_d) \Big], 
\end{equation}
Since we have only added a gauge singlet the $b,\, B$ and $d$ coefficients are unaltered.  The new coefficient is,
\begin{equation}
d_B = \Big( \frac 3 {20}, \frac 1 4 ,0\Big)~. 
\end{equation}
In turn these new couplings have their own RGEs,
\begin{eqnarray}
\kappa\, \beta^{(1)}_{\gu} &=& \frac 5 4 \gu^3 + \gu\,
\Big( 2\,\gd^2 + 3\,y^2_t - \Big( \frac 9 4 g^2_2 + \frac{9}{20}\,g^2_1 \Big) \Big)~,
 \\
\kappa^2\, \beta^{(2)}_{\gu} &=& -\frac 3 4 \gu^5 + \gu^3
  \Big(-\frac{27}{8}\, y^2_t + \frac{165}{32}g^2_2 
       + \frac{309}{160}g^2_1 - 3\,\lambda_h - \frac{15}{4}\gd^2 \Big) \nonumber \\ 
& &{} + \gu \Big(-\frac{27}{4} y^4_t + y^2_t \Big( 20\, g^2_3 
  + \frac{17}{8}g^2_1  + \frac{45}{8} g^2_2 \Big) 
  - \frac{21}{4}g^4_2 - \frac{27}{20\,} g^2_2\,g^2_1 + \frac{117}{200}\,g^4_1 
  + \frac 3 2 \lambda_h^2 \nonumber \\
& &{} + \gd^2\Big( \frac{39}{8}\,g^2_2 + \frac{3}{40}\, g^2_1- 3\lambda_h\Big)
      - \frac{21}{4} \gd^2\, y^2_t  - \frac 9 4 \gd^4\Big)~,
\end{eqnarray}
and
\begin{eqnarray}
\kappa\, \beta^{(1)}_{\gd} &=& \frac 5 4 \gd^3 + \gd\,
\Big( 2\,\gu^2 + 3\,y^2_t - \Big( \frac 9 4 g^2_2 + \frac{9}{20}\,g^2_1 \Big) \Big)~,
\\
\kappa^2\, \beta^{(2)}_{\gd} &=& - \frac 3 4 \gd^5 + \gd^3\,
  \Big( -\frac{27}{8}\, y^2_t + \frac{165}{32}g^2_2 + \frac{309}{160}g^2_1 
  - 3\,\lambda_h - \frac{15}{4}\gu^2  \Big) \nonumber \\ 
& &{} + \gd \Big(-\frac{27}{4} y^4_t + y^2_t 
      \Big( 20\, g^2_3 + \frac{17}{8}g^2_1  + \frac{45}{8} g^2_2 \Big) 
      - \frac{21}{4}g^4_2 - \frac{27}{20\,} g^2_2\,g^2_1 + \frac{117}{200}\,g^4_1 
      + \frac 3 2 \lambda_h^2 \nonumber \\
& &{} + \gu^2 \Big( \frac{39}{8}\,g^2_2 + \frac{3}{40}\, g^2_1 - 3\lambda_h\Big) 
      - \frac{21}{4} \gu^2\, y^2_t  - \frac 9 4 \gu^4\Big)~.
\end{eqnarray}
These new couplings also enter in the running of the quartic and the top Yukawa. 
These top Yukawa RGE is given by,
\begin{eqnarray}
\kappa\, \beta^{(1)}_{y_t} &=& \frac 9 2 y^3_t 
  + y_t \Big( \frac 1 2  \gu^2  + \frac 1 2 \gd^2 
              - \Big( 8\, g^2_3 + \frac 9 4 g^2_2 + \frac{17}{20}\, g^2_1 \Big)\Big)~,
\\
\kappa^2\, \beta^{(2)}_{y_t} &=& -12\,y^5_t + y^3_t 
  \Big( 36\,g^2_3 + \frac{225}{16}g^2_2 + \frac{393}{80}\,g^2_1 - 6\,\lambda_h
        - \frac{9}{8}( \gu^2 + \gd^2) \Big) \nonumber \\
& &{} + y_t\,\Big( -108\, g^4_3 + 9\, g^2_3 g^2_2 + \frac{19}{15}g^2_3\,g^2_1
      - \frac{21}{4} g^4_2 - \frac{9}{20} g^2_2\, g^2_1 
      + \frac{1303}{600}\, g^4_1 +  \frac 3 2\, \lambda_h^2 \nonumber \\
& &{} +  \Big(\frac{15}{16} g^2_2+ \frac{3}{16} g^2_1 \Big)\,(\gu^2 + \gd^2) 
      - \frac{9}{16}(\gu^4 + \gd^4) - \frac 5 4 \gu^2\gu^2 \Big)~. 
\end{eqnarray}
The quartic RGE is,
\begin{eqnarray}
\kappa\, \beta^{(1)}_{\lambda_h} &=& 12\, \lambda_h^2 
  + \lambda_h \Big(12\, y^2_t - 9\, g^2_2 - \frac 9 5\, g^2_1 + 2 (\gu^2 + \gd^2) \Big)
  - 12\, y^4_t \nonumber \\
& &{} + \frac 9 4 g^4_2 + \frac{9}{10} g^2_2 g^2_1 + \frac{27}{100}\, g^4_1 
      - (\gu^2 + \gd^2)^2~, \\
\kappa^2\, \beta^{(2)}_{\lambda_h} &=& -78\, \lambda_h^3 + \lambda_h^2 
  \Big( 54\, g^2_2 + \frac{54}{5}\, g^2_1 -72\, y^2_t - 12\, (\gu^2 + \gd^2) \Big) 
  \nonumber \\
& &{} + \lambda_h\, \bigg( -3\, y^4_t + 80\, g^2_3\, y^2_t 
      + \frac{45}{2} g^2_2\,y^2_t + \frac{17}{2} g^2_1\,y^2_t 
      - \frac{33}{8}\,g^4_2 + \frac{117}{20}\, g^2_2\,g^2_1 
      + \frac{2007}{200}\,g^4_1 
      \nonumber \\ 
& &{} \qquad + \Big( \frac{15}{4} g^2_2 + \frac 3 4 g^2_1 \Big) (\gd^2 +  \gu^2) 
      - \frac 1 4 \gu^4 - \frac 1 4 \gd^4 + 3\,\gu^2 \,\gd^2 \bigg)
      \nonumber \\
& &{} + 60\, y^6_t - y^4_t \Big( 64\, g^2_3 + \frac{16}{5}\, g^2_1\Big) 
      + y^2_t \Big(-\frac{9}{2}\, g^4_2 
      + \frac{63}{5}g^2_2\, g^2_1- \frac{171}{50}\,g^4_1 \Big) 
\nonumber \\
& &{} + \frac{273}{8} g^6_2 - \frac{321}{40} g^4_2\, g^2_1
      -\frac{1773}{200}g^2_2\,g^4_1
      - \frac{3699}{1000} g^6_1 \nonumber \\
& &{} - \Big( \frac 3 4 g^4_2 + \frac{3}{10}\, g^2_2\, g^2_1
+  \frac{9}{100}\,g^4_1 \Big)(\gu^2 + \gd^2) + \frac 5 2 (\gu^6 + \gd^6) 
+ \frac{17}{2}\gu^2\,\gd^2\,(\gu^2 + \gd^2) \, . \;\;\qquad
\end{eqnarray}

\subsection*{The MSSM with Adjoints}

The final epoch we are interested in occurs in both Model I and II once all the superpartners, the second Higgs doublet, and the adjoint chiral super fields are included.  The field content is that of the MSSM with additional adjoint fermions and scalars.  As the adjoints have no supersymmetric interactions outside of the kinetic term, no new couplings are introduced and all $O(g^3 y^2)$ pieces of the gauge couplings RGEs are the same as in the MSSM.  Namely the coefficients in (\ref{eq:2loopRGE}) are,
\begin{equation}
b = \Big( \frac{33}{5}, 3, 0 \Big)\, \quad B = \left( \begin{array}{ccc} \frac {199}{25} & \frac{27}{5} & \frac{88}{5} \\ \frac 9 5 & 49 & 24 \\ \frac{11}{5} & 9 & 68 \\ \end{array}\right)\, \quad d = \Big(\frac{26}{5}, 6,\ 4 \Big).
\end{equation}
Since we are now in a supersymmetric theory the Higgs quartic is no longer a separate coupling but is instead determined from the D-terms in terms of gauge couplings.  This leaves only the top Yukawa, which runs as
\begin{eqnarray}
\kappa\, \beta^{(1)}_{y_t} &=& 6y_t^3 -y_t
 \Big(\frac{16}{3}g_3^2+3g_2^2+\frac{13}{15}g_1^2 \Big) \\
\kappa^2\, \beta^{(2)}_{y_t} &=& 
  -22 y_t^5 + y_t^3 \Big(16g_3^2 +6g_2^2+\frac{6}{5}g_1^2\Big) \nonumber \\
& &{} + y_t \Big(-\frac{16}{9}g_3^4 + 8g_2^2 g_3^2 
+ \frac{136}{45}g_1^2 g_3^2 
+ \frac{15}{2}g_2^4 + g_1^2 g_2^2 
+ \frac{2743}{450}g_1^4\Big)~.
\end{eqnarray}


\end{document}